\documentclass[a4paper, 11pt]{article}

\usepackage{dcolumn}% Align table columns on decimal point
\usepackage{bm}% bold math

\usepackage[utf8]{inputenc}
\usepackage{mathptmx}   % Set Font to Times
\usepackage{tabularx}
\usepackage[margin=2cm]{geometry}
\usepackage{csquotes}
\usepackage[usenames,dvipsnames]{color}

\usepackage{multirow}
\usepackage{array}
\usepackage{graphicx}
\usepackage{overpic}
\usepackage{subfigure}
\usepackage{eurosym}
\newcolumntype{L}[1]{>{\raggedright\let\newline\\\arraybackslash\hspace{0pt}}m{#1}}
\newcolumntype{C}[1]{>{\centering\let\newline\\\arraybackslash\hspace{0pt}}m{#1}}
\newcolumntype{R}[1]{>{\raggedleft\let\newline\\\arraybackslash\hspace{0pt}}m{#1}}

\usepackage[square,numbers]{natbib}
\usepackage{amsmath,amssymb}
\usepackage[normalem]{ulem}
\usepackage{color}
\usepackage{bbding}
\usepackage{url}
\usepackage{placeins}

\definecolor{ao}{rgb}{0.0, 0.5, 0.0}

\begin{document}

\title{Viscous rebound of a quasi-2D cylinder on a solid wall}

\author{Alicia Aguilar-Corona$^1$ ,Micheline Abbas$^{2,*}$,Matthieu J. Mercier$^3$, Laurent Lacaze$^3$.\\\\
$^1$Facultad de Ingenieria Mecanica, Universidad Michoacana de San Nicolas de Hidalgo,\\CP 58030, Morelia, Michoacan, Mexico\\
$^2$Laboratoire de Génie Chimique, Université de Toulouse, CNRS, INPT, UPS, Toulouse, France\\
$^3$Institut de Mécanique des Fluides de Toulouse (IMFT), Université de Toulouse,\\
CNRS, Toulouse, France\\\\
$^*$micheline.abbas@ensiacet.fr}

\date{\today}% It is always \today, today,

\maketitle

\section{Abstract}

The purpose of the present study is to extend the simple concept of apparent coefficient of restitution, widely approached in the literature for the case of a single-contact-point between a sphere and a wall, to the case of bouncing whose complexity is increased due to the shape of the contacting object. %the influence of an apparent roughness of the solid surfaces on the threshold of bouncing, as already developed in the literature, to the influence of an apparent softness on the contact time beyond only dissipation, and encompassing both liquid and solid responses during contact.
For this purpose, experiments are carried out with a finite-length cylinder, freely falling in a liquid at rest. A rigid tail is attached to the cylinder, allowing to maintain a vertical trajectory and to keep the axis of the cylinder parallel to the bottom wall down to small gap between them. Yet, more complex 3D motions of the cylinder with respect to the wall occur during bouncing, including multi-contact-points between the cylinder and the bottom as well as cavitation. When the Stokes number $St$ (the ratio between characteristic inertial forces experienced by the particle compared to viscous forces in the fluid) is increased, the experimental results suggest that the ratio of the apparent coefficient of restitution to the solid one increases from $0$ (at low $St$) to 1 (at large $St$) with a critical Stokes number $St_c$ below which no bouncing is observed, as usually obtained in the literature for a single-contact-point bouncing sphere.
%It is found here that \textcolor{ao}{$St_c\approx 75$} is larger than that of a sphere ($St_c\approx 10$) for comparable surface roughness.
%In the transition regime, the coefficient of restitution increases from $0$ to $1$ along approximately two decades in $St$, with a significant dispersion. 
In a conceptual approach of the influence of a cut-off length scale prior contact and a contact time scale on the observed experimental results, we investigated numerical modelling of an idealized situation (2D infinite cylinder falling parallel to the wall). To this end, we carried numerical 2D simulations where the fluid equations of motion were coupled to the particle equation of motion through an Immersed Boundary Method. The particle equation of motion was coupled to an elastic force to model bouncing. {This numerical model requires parameterization of the cut-off length (interpreted as a roughness) and the contact time (associated with the contact elasticity) used here to capture the experimental observations.} The simulations confirmed that i) the departure of the coefficient of restitution from 0 is strictly dependent on the apparent roughness and ii) the coefficient of restitution depends on the contact time. Finally, in an effort to rationalize the experiments and the simulations on such conceptual approach, we model the coefficient of restitution as the product of two contributions to the mechanical loss of energy: the collision-to-terminal velocity ratio ($V_c/V_t$) of the approach-phase and the rebound-to-collision velocity ratio ($-V_r/V_c$) of the contact-phase. We then interpreted the experimental measurements in light of this model, showing evidence that the assumption of a global relationship between the contact time and an apparent roughness (all being linked to the bouncing complexity including multi-contact-points and cavitation in experiments) leads to a reasonably good prediction of the coefficient of restitution in the intermediate regime in $St$. This suggests the relevance of lumping the complex details of physical phenomena involved during contact into a simple concept based on the contact apparent roughness and elasticity.

\section{Introduction}
\label{sec:intro}

The interaction of solid particles evolving in a viscous fluid has been considered in many studies for its obvious relevance in many geophysical and industrial applications involving multiphase flows. The related key questions are numerous ranging from the mesoscale dynamics, that is the collective motion of the suspended particles, to the microscale dynamics at the scale of the grain. In particular, if one simplifies such complex systems by considering only two approaching particles, one handles local processes such as dissipation induced by their interaction. This includes the viscous dissipation due to fluid motion induced by their relative motion close to contact as well as the mechanical dissipation during solid contact. 

A main issue, when dealing with such apparently simple systems, is the antagonism between geometrical and mechanical properties of modelled particles and the ones of real particles. Then, two main aspects are usually modelled, as opposed to solved, which are (i) the local deformation due to elasticity of the particles and (ii) the surface roughness. Otherwise, infinitely rigid and smooth objects would lead to a complete dissipation of the initial kinetic energy, and no bouncing could occur. Such dissipation would be due to the lubrication force induced by the viscous flow in the gap between particles diverging when the gap goes to zero for perfectly smooth and rigid particles as evidenced by\cite{Brenner1961}in the case of a sphere approaching a wall. 
However, a full dissipation preventing bouncing has been shown through several experimental studies to be unlikely, at least above some given threshold of inertia close to contact \citep{Barnocky1988,Joseph2001,Gondret2002,Chastel2016,Birwa2018,Mongruel2020}. Therefore, the singularity of the lubrication force at zero gap, that would only lead to solid contact on a infinite time scale with zero relative velocity, should somehow be regularised. This is why modelling the bouncing of a sphere is often  sought as a regularization of the physical system.

The elastohydrodynamic collision model proposed by \cite{Davis1986} and \cite{Lian1996} is a regularization approach incorporating (i), which thus allows bouncing while preventing solid contact. This led to several experimental studies dealing with the bouncing of spherical particle on a rigid wall, for which the surface roughness of particle could not be disregarded \citep{Barnocky1988}. Then, on the opposite side, the possibility of rigid grain to bounce only due to surface roughness has been addressed in the literature \citep{Chastel2016}. As suggested by \cite{Cawthorn2010},  (ii) proving solid contact required sharp shape roughness, otherwise the elastohydrodynamic collision is only scaled down to the roughness scale. Trying to separate (i) from (ii), \cite{Birwa2018} has experimentally highlighted the possibility of solid contact between rough rigid objects on a finite time scale. An extension of the elastohydrodynamic collision including (ii) could therefore be of relevance \citep{Yang2008}. Note that additional aspects can be also found in the literature, as for instance the liquid-glass transition in the presence of highly viscous liquid films between colliding surfaces \cite{donahue2012agglomeration, davis2019simultaneous}.

From a numerical modelling point of view, the scale of regularization, either due to (i) or (ii), is usually too small to be reasonably resolved, in particular for complex systems involving several particles. Actually, significant advances have been recently made on the numerical modelling of complex shape and non-rigid particles \citep[see for instance][]{Mollon2016}. However, these methods still suffer from a limitation due to the computational cost and the spatial scale of shape complexity, which usually lead to locally smooth surface. Accounting for the shape complexity of a body does not necessarily allow one to capture the required change of scale of surface roughness that would be at the origin of the regularization at contact mentioned previously. Then, a specific attention is still required to incorporate the small-scale physical processes as closure model into classical numerical approach such as Discrete Element Method (DEM), for which (i) and (ii) are not fully resolved. This approach has been extensively proposed in the literature using a coupled DEM-DNS solver for the solid-fluid resolutions \citep{Ardekani2008,Yang2008,Feng2010,Li2012,Kempe2012,Brandle2013,izard2014modelling,Li2020,wachs2023modeling}. All these approaches are somehow similar, as the meshgrid of the fluid solver covers the entire domain including the solid object to prevent from complex and costly numerical algorithm when dealing with body-fitted grid methods, which would moreover be incompatible with an objective multi-body systems. For these fixed-grid methods, an extra term is required to enforce the solution within solid body. Methods can slightly differ but mostly lead to the same conceptual methodology. In this case, the above mentioned singularity, or lack of resolution, occurs at the meshgrid of the fluid solver, when the two solid surfaces close to contact reach a distance of the order of the mesh-size. Due to numerical limitation, this length scale can actually be quite important compared to the typical diameter of the particles $D$, i.e. only one order of magnitude smaller $D/10$. As this scale is often thought as large compared with the roughness of real grains, a subgrid lubrication model is often added when the solids get closer. Moreover, a separation between solid contact time scale and fluid time scale is often considered, assuming strict rigidity of the solids. These two aspects (lubrication model and rigidity) still require some attention. In particular, adding a lubrication model allows to delay the singularity to a smaller scale, and therefore to obtain a better quantitative agreement with experiments \citep[as the fluid dissipation is increased to a level closer to real situation, see][for instance]{izard2014modelling}. However, this remains quite an empirical alternative. Using Immersed Boundary Method (IBM) on a fixed grid, \cite{Li2020} attempted to refine the meshgrid in the gap between solid object (sphere/wall) down to $\sim D/1000$ using local refinement and taking advantage of geometrical symmetry. Doing so, a subscale lubrication model is not added, allowing to capture most of the dissipation as the mesh reaches a length scale comparable to real roughness. Then, roughness is seen as an apparent roughness based on the meshgrid \citep[see supplementary material in ][]{Lacaze2021}. Even if the link between numerical dissipation and real dissipation remains unclear, the scale issue requiring an additional lubrication model is at least removed with such approach.    

Quite surprisingly, whatever the `philosophy' of regularization or physical process at the microscopic scale, a robust observation obtained from experiments and modelling, either numerical or theoretical, is the evolution of an effective coefficient of restitution including both viscous dissipation and mechanical/structural elasticity, with the dimensionless Stokes number $St$, measuring the ratio of particle inertia to fluid dissipation as explicitly defined in eq. (\ref{eq:St}). In particular, a critical Stokes number $St_c$ is obtained delineating a viscous region for which no bouncing is observed ($St<St_c$) and an inertial region where bouncing occurs ($St>St_c$). This robustness allowed numerical simulations, as discussed later on, to be relevant to predict this the transition from viscous to inertial regimes, even if the local mechanisms allowing this bouncing (i vs ii) remain unclear. Then, it would suggest that any complex mechanism at play during bouncing, due either to the local complex shape of the bouncing particle or to deformation and/or to phase change (in the liquid) induced by strong local pressure just prior contact and/or low local pressure upon motion reversal, respectively, can be captured by simplified modelling based on effective bouncing characteristics. The terms elasticity and roughness, classically used in the literature of IBM/DEM modelling, thus refer to the adjustable bouncing parameters which are actually the time scale of contact and the regularization length prior to contact, to be parameterized in an IBM/DEM bouncing simulation. Then even if discriminating and modelling elasticity versus roughness remain uncertain, their conceptual consequences on the effective bouncing are promising and still deserve attention. One proposes to adopt this approach to extend the bouncing models developed for the case of a sphere \citep{izard2014modelling} towards a conceptual approach of apparent elasticity and apparent roughness as a generalization of more complex bouncing configurations. 

Before ending this section, we define the Stokes number of an object falling under gravity in a viscous fluid. This dimensionless number is intimately related to the settling Reynolds number, defined here as
\begin{equation}
Re = \frac{\rho_f D V_t}{\mu} ,
\label{eq:Re}
\end{equation}
The Stokes number is then built from the ratio between the object relaxation time and the fluid motion at the scale of the object. Instead of considering different expressions associated with the cylinder settling in the experiments and numerical simulations (since the shape is not exactly the same), we use for simplicity the commonly used expression for a falling sphere: 
\begin{equation}
St = \frac{1}{9} \frac{\rho^* D V_t}{\mu} = \frac{1}{9} \left(1+\frac{\rho_f}{\rho_p} C_M \right) \frac{\rho_p}{\rho_f}  Re.
\label{eq:St}
\end{equation}
Here $\rho_f$ and $\mu$ correspond to the fluid density and viscosity. $D=2R$ indicates the cylinder diameter, $R$ its radius. $\rho^*= \rho_p\left(1+\frac{\rho_f}{\rho_p} C_M \right)$ is the effective cylinder density taking into account the added mass effect and $C_M$ represents the added mass coefficient. $V_t$ corresponds to the cylinder terminal settling velocity, reached when the drag force balances the cylinder apparent weight. 

\textbf{\textit{Paper structure:}}

\noindent Experiments (section \ref{sec:exp}) and simulations (section \ref{sec:Sim}) are used to investigate the coefficient of restitution $e$ of a cylinder bouncing on a `smooth' wall, as a function of its inertia characterized by the Stokes number. Experiments are designed to provide the bouncing of a finite-length cylinder onto a horizontal surface while simulations are designed to provide a twin 2D situation of a 2D cylinder bouncing along a horizontal line, in which roughness and elasticity are therefore subscale models which could mimic the complexity of finite-length 3D bouncing. Despite the apparent simplicity of the experimental setup, it is shown that the values of the coefficient of restitution as a function of the Stokes number are not easily rationalized onto a master curve, but a shift remains between the reported curves. In order to interpret the data, we postulate that i) the scatter is mainly associated with the cylinder pitching at the onset of the collision process and the generation of cavitation bubble at contact, and that ii) the physical phenomena at play during the contact can be interpreted in terms of a cut-off length at contact and a contact time. In the numerical simulations, as the configuration remains 2D all along the particle trajectory (collision along a contact line), $\eta$ appears to influence mainly the value of the critical Stokes number, while the contact elasticity parameter $\alpha$ (that allows to tune the characteristic collision time) is shown to lead to different levels of the $e-St$ curves. 
Section \ref{sec:model} discusses a general reduced-order model for the restitution coefficient, that involves a dimensionless apparent roughness parameter $\eta$ (which sets a critical Stokes number above which rebound occurs) and where the contact elasticity is expressed in terms of a finite contact time scale (which decreases when the Stokes number increases).
The paper ends with a discussion section \ref{sec:discussion}, where we show that using the numerically and experimentally measured contact time, we capture the different levels of $e-St$ curves in a reasonable way. This confirms the importance of measuring contact time along the process of characterization of the complex collision process.

%%%%%%%%%%%%%%%%%%%%%%%%%%%%%%%%%%%%%%%%%%%%%
\section{3D bouncing: experiments}
\label{sec:exp}

\subsection{Experimental setup}
\label{sec:exp_setup}

To model an idealized rebound in a laboratory setup, we consider the controlled settling of a finite width cylinder of diameter $D$ and length $L$ (larger than D), being constrained to fall between two vertical walls separated by a distance $L(1+\epsilon)$ with $\epsilon$ very small compared to $1$. This setup is aimed to be designed to be close enough to a 2D cylinder bouncing as characterised in the previous section, anticipating an effective 2D description of the experimental rebound. For this purpose, the motion of the cylinder is imposed to take place in a plane perpendicular to its long axis and along the vertical direction. In practice, three-dimensional motion are unavoidable, due to both the finite length of the cylinder and wake instabilities in the vertical plane. These three-dimensional motions require to be controlled enough. For this purpose, the gap between the cylinder and the lateral walls (L$\epsilon$) is always sufficient to allow the cylinder falling but small enough to reduce three-dimensional rotational to a range of small angles (typically $L\epsilon\simeq0.1$\,cm). 
 Moreover, to minimize three-dimensional motion, the cylinder is provided with a long thin tail of dimension $\ell\gg D$ and with a similar length $L$. 
 \begin{figure}
  \centering
  \hspace*{0cm}(a) \hspace{4cm} (b) \hspace{3.5cm} (c)\hspace{3.5cm} (d)\\
  \includegraphics[width=0.95\linewidth]{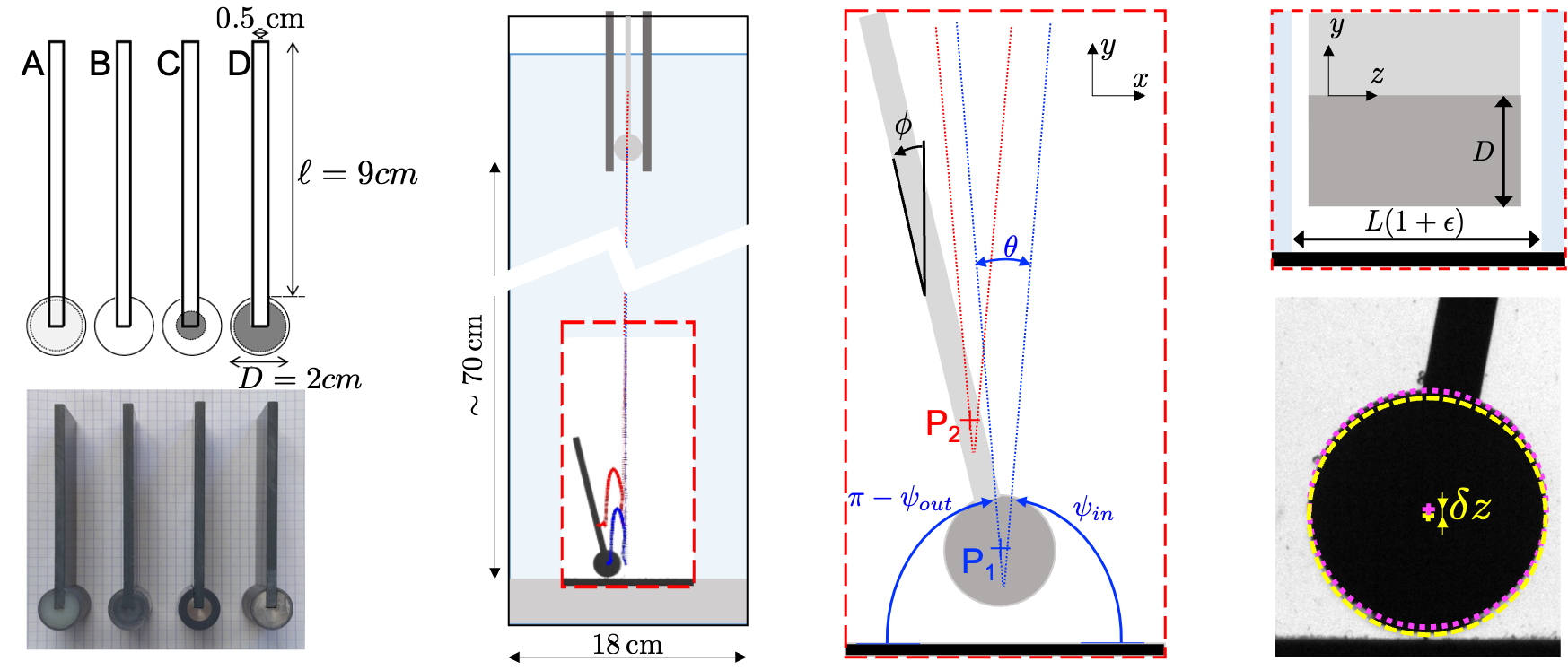}\\
  \vspace{-4.3cm}
  \hspace{13.1cm} (e)\\
  \vspace*{3.4cm}
  \caption{(a) Design and picture of the four cylinders used. (b) Schematic of the experiment, the delimited area in red is the field of view of the camera. Schematic and definitions of the extracted quantities in the (c) $(x-y)$ plane and (d) $(y-z)$ plane, and (e) from real images.} \label{fig:exp-setup}
\end{figure}
 
 In order to vary the Stokes number $St$ in the experiments, several cylinders and fluids have been used. The different cylinders are shown in Fig.\,\ref{fig:exp-setup}(a), all of them having an outer core of PVC but their inner part can be replaced by another material to modify the object density but not its surface/contact properties. The tail is a plate made of PVC as well.
The cylinder properties are summarized in table \ref{tab:cylinders}.
The rebound of the cylinders have been studied in four different fluids at room temperature ($\sim20\deg$): air, salt water, and a mixture of salt water with Ucon oil (DOW\copyright) 75H900000 at $10\%$ and $15\%$ in mass. Their densities and kinematic viscosities are summarized in table \ref{tab:fluids}.

\begin{table}[htb]
    \centering
    \begin{tabular}{l|c|c|c|c}
         & A & B & C & D\\
         \hline
         inner  material & PMMA & PVC & Steel & Steel\\
         inner  diameter (cm)& 1.9 & N.A. & 1.0 & 1.8\\
        $D$ (cm) & 2.074 & 2.071 & 2.072 & 2.072\\
        $L$ (cm) & 3.95 & 3.96 & 3.91 & 3.96\\
        $\ell$ (cm) & 9.0 & 9.0 & 9.0 & 9.0\\
        %Mass (g) & 42.1 & 44.1 & 66.6 & 98.5\\
        $\rho_p$ (g/cm$^3$) & 1.36 & 1.43 & 2.17 & 3.19\\
    \end{tabular}
    \caption{Properties of the cylinders shown in Figure \ref{fig:exp-setup}(a).}
    \label{tab:cylinders}
\end{table}
\begin{table}[htb]
    \centering
    \begin{tabular}{c|c|c}
         fluid & $\rho$ (kg/m$^3$) & $\nu$ ($10^{-6}$ m$^2$/s) \\
         \hline
         air & 1.17  & 15.7\\
         salt water & 1050 & 1.1\\
         ucon/salt water ($10/90$)& 1050 & 8.4\\
         ucon/salt water ($15/85$)& 1050 & 17.7\\
    \end{tabular}
    \caption{Properties of the fluids used. Mixture ratio are percentage in mass.}
    \label{tab:fluids}
\end{table}

As shown in Fig.\,\ref{fig:exp-setup}(b), the glass tank horizontal dimensions are $18.0\times4.0$\,cm$^2$ and its vertical extent of $74.0$\,cm allows for a falling distance of approximately $70$\,cm. The cylinder, almost fully immersed, is released at the top of the tank by opening a gate (fast mechanical aperture using a compressed air piston) and falls due to gravity in a fluid at rest until reaching the bottom of the tank where it encounters a fixed flat surface made of PVC (same material as the outer part of the cylinder). We follow its dynamics with a high-speed camera connected to a telecentric lens to avoid parallax effects, the sampling frequency chosen is 2kHz and the field of view is $1000\times2000$ px$^2$ with a resolution of $\sim0.1$mm/px observing the $(x,y)$ plane as shown in Fig. \ref{fig:exp-setup}(b,c). The gray levels due to the back-lightning of the tank allows to precisely extract the contour of the object, and even its orientation with respect to the $(x,y)$ plane. A series of experiments have also been investigated using a Phantom\copyright high-speed camera, with the sampling frequency at 130kHz (in salt water only).

For each grey-level image, we extract the contour and the center ($P_1$) of the cylinder, as well as the contour and the barycenter ($P_2$) of the whole object. From the temporal evolution of those two points, we can follow the orientation in the $(x,y)$-plane of the tail $\phi(t)$ which corresponds to the angle of the line $(P_1,P_2)$ with the vertical; the velocity for the trajectories of $P_1$ and $P_2$ and their angle with the horizontal are found by computing a linear fit of the trajectory over $\sim20$ time-steps.
We define more specifically $\psi_{in}$ and $\psi_{out}$ the angles of the trajectory of the center ($P_1$) with the horizontal, just before and after rebound, and $\theta=|\psi_{out}-\psi_{in}|$ the difference between these two angles.
The coefficient of restitution in the vertical direction, denoted $e_y$ in the following, will be computed by making the ratio of the velocity of $P_1$ just before and after contact.
As illustrated in  Fig.\ref{fig:exp-setup}(d), by using the gray-level to discriminate the front (black) and rear (lighter gray) faces of the cylinder, we can extract the difference in height $\delta z$ of their centers which relates to the orientation out-of-plane of the cylinder ($\sin^{-1}(\delta z/L)$). % indicating the level of "roughness" in experiments.

The release of the cylinder is always done in a fluid at rest for several minutes, with the initial position of its tail as vertical as possible. Nevertheless, residual perturbations in the fluid or at release can generate a small rotation of the tail angle with the vertical ($\phi$), which can stay almost constant in viscous fluids, but might generate lift and some rotational motions in general. A precise monitoring of all these angles are done for each experiment.

\subsection{Results}
\label{sec:exp_results}

\subsubsection{Contact time and out-of-plane inclination at rebound}

The rebound on the bottom wall observed with this experimental setup was shown to be nearly 2D. However, a deeper investigation indicates a slightly more complex bouncing feature, highlighting 3D contribution in the out-of-plane direction. The complexity of the bouncing is associated with the out-of-plane orientation of the cylinder with respect to the horizontal, as discussed in the previous section and quantified by the length $\delta z$. This leads to a contact point of the cylinder (either front or rear section touching the bottom plate) instead of an idealized contact line along the entire cylinder, as would be obtained for a purely 2D bouncing. The full bouncing process of such apparent 2D bouncing can actually be characterized by a succession of bouncing of the front and rear sections of the cylinder on a very short time scale, prior to a significant take off of the object indicating the end of the apparent 2D bouncing. Such complex bouncing can be classified by the number of visible contacts between the cylinder and the plate. It can have one (front or rear part depending on its out-of-plane inclination), two (front and rear) or even three (front/rear/front or vice-versa) contacts with the PVC plate before going away from it.
Accordingly, we can also define an apparent contact time $t_C$ that corresponds to the delay between the first and last images showing contact, with a lower bound being the frame rate of the camera ($0.5$ms) for the $1$-contact cases.
Two examples are illustrated in Fig.\,\ref{fig:timelapse}(a,b), corresponding movies are given in supplementary material\,\footnote{WATER\_CYLA\_1CONTACT\_2000FPS.gif, WATER\_CYLD\_3CONTACTS\_2000FPS.gif, and WATER\_CYLD\_1CONTACT\_CAVITATION\_130000FPS.gif}.
\begin{figure}
\centering
\begin{overpic}[trim={0 0 0 0}, clip=true,width=0.95\textwidth]{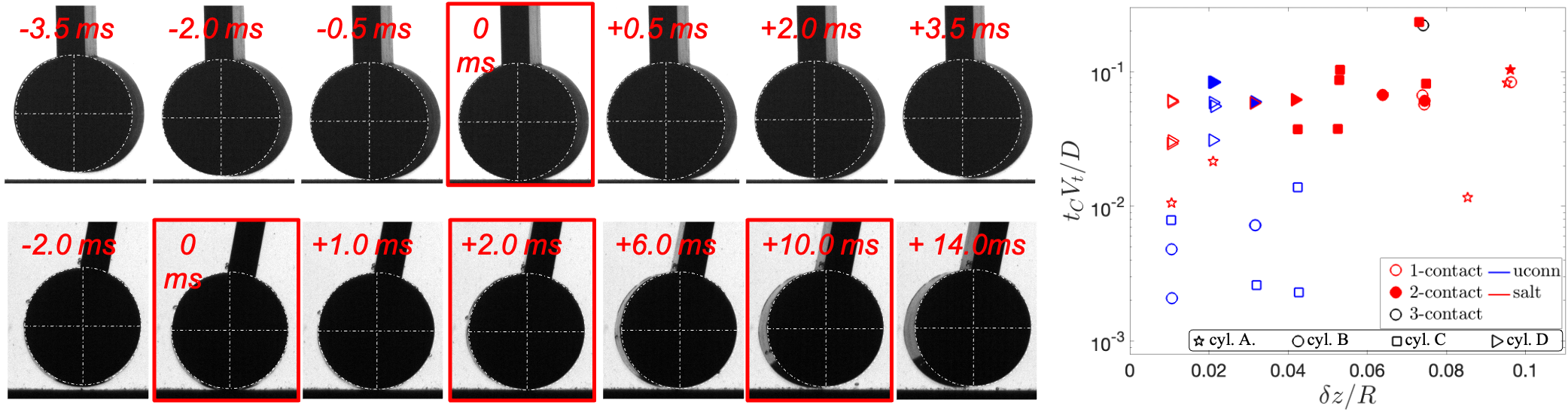}
\put(-2,24){(a)}
\put(-2,10){(b)}
\put(67,24){(c)}
\end{overpic}
\vspace{-0.1cm}
    % (a)\\
    % \includegraphics[width=0.95\linewidth]{MA_E3_1contact.png}\\
    % (b)\\
    % \includegraphics[width=0.95\linewidth]{PVC_E5_3contacts.png}
    % \includegraphics[width=0.8\linewidth]{all_contacts_v3.png}\\
    \caption{Series of images to illustrate rebounds with (a) 1-contact (front contact) for cylinder A in salt water, and (b) 3-contact (rear/front/rear front contact) case for cylinder B in salt water. The frames with contacts are indicated in red, a dashed-dot white circle shows the front face of the cylinder.
    (c) Sorting of rebound cases based on $\delta z/R$, the non-dimensional contact time and number of contacts for cases in uconn/salt water (blue) and salt water (red). Cylinders are sorted by symbols.}
    \label{fig:timelapse}
\end{figure}
%\begin{figure}[tb]
%\centering
%\begin{overpic}[trim={0 0 0 0}, clip=true,width=0.8\textwidth]{all_contacts_v4.png}
% \begin{overpic}[width=0.8\textwidth,grid,tics=10]{all_contacts_v3.png}
%\put(-3,80){(a)}
%\put(-3,58){(b)}
%\put(-3,38){(c)}
%\put(52,38){(d)}
%\end{overpic}
%\vspace{-0.1cm}
    % (a)\\
    % \includegraphics[width=0.95\linewidth]{MA_E3_1contact.png}\\
    % (b)\\
    % \includegraphics[width=0.95\linewidth]{PVC_E5_3contacts.png}
    % \includegraphics[width=0.8\linewidth]{all_contacts_v3.png}\\
%    \caption{Series of images to illustrate rebounds with (a) 1-contact (front contact) for cylinder A in salt water, and (b) 3-contact (rear/front/rear front contact) case for cylinder B in salt water. The frames with contacts are indicated in red, a dashed-dot white circle shows the front face of the cylinder.(c,d) Sorting of rebound cases based on $\delta z/R$, the time and number of contacts for cases in uconn/salt water (blue) and salt water (red). [MAT- data in air to add to panel (c)?]}
 %   \label{fig:timelapse}
%\end{figure}
%
We have also registered the out-of-plane inclination at the rebound, or equivalently $\delta z/R$, to relate it to the nature of the rebound. On average, image analysis led to estimates of $\delta z/R$ smaller than $0.1$, which corresponds to inclinations smaller than $5\deg$.
{Even if the non-dimensional apparent contact time $t_C V_t /D$ roughly increases with $\delta z/R$, dispersion of the results does not allow to assure that they are strongly correlated, as illustrated in Fig.~\ref{fig:timelapse}(c). One can notice however that it is more likely to have two or three contacts at rebound when the out-of-plane inclination is important, i.e. for increasing $\delta z/R$.}

\subsubsection{Influence of cavitation at rebound}

\begin{figure}
    \begin{overpic}[trim={0 0 0 0}, clip=true,width=0.95\textwidth]{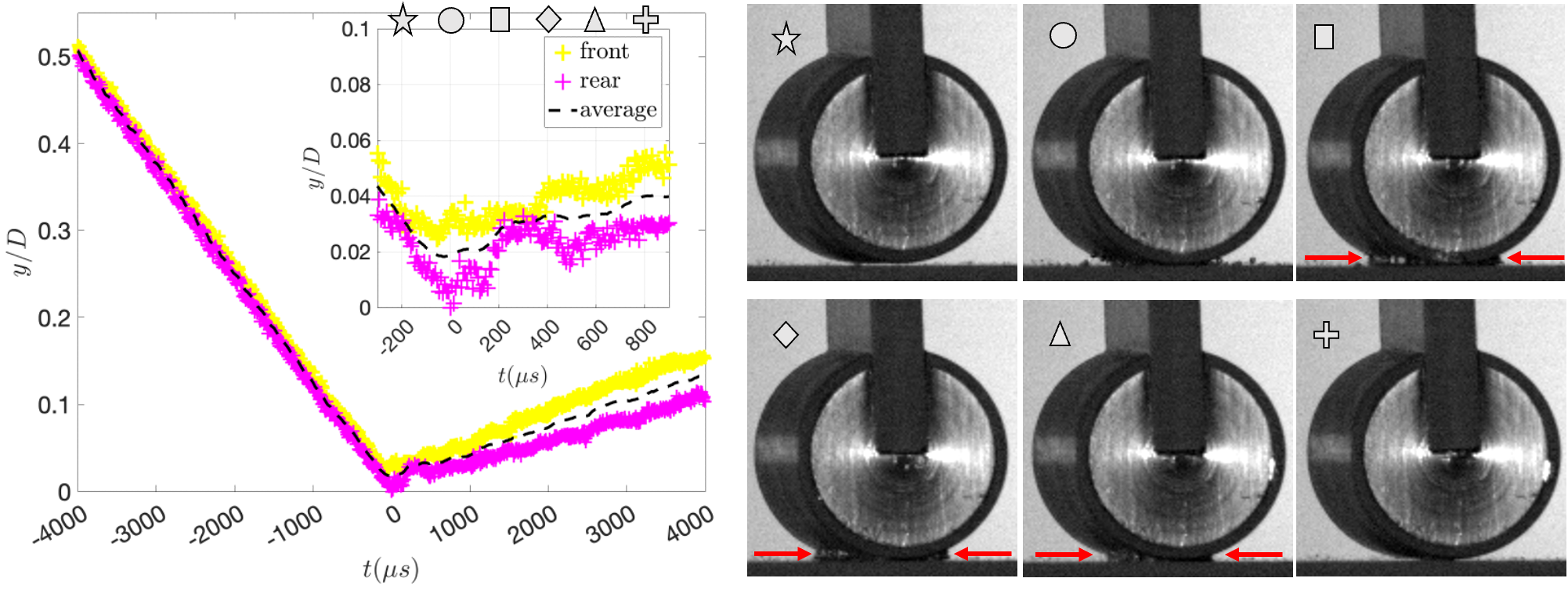}
    \put(0,36){(a)}
    \put(46.5,36){(b)}
    \end{overpic}
    \caption{Cylinder D in water. (a) Position of the front and rear faces of the cylinder (as well as the average of the two) with time, with the inset zooming on the $1 ms$ around the rebound. (b) Series of images from the camera (every $200\mu s$) with symbols indicating the corresponding times in the inset in (a), the red arrow indicates the cavitation bubble when visible.}
    \label{fig:cavitation}
\end{figure}

For some of our experiments, the estimated values for $t_C$ is at the limit of the sampling frequency of the camera (2kHz). In order to better resolve those values, a series of experiments have been repeated with a camera working at very high-speed (130kHz). 
As can be seen in Fig.\,\ref{fig:cavitation}(a) for a cylinder (type D) in water, the overall vertical dynamics is well represented by two branches of trajectories at constant speed. although some out-of-plane inclination is visible after the rebound. If we focus on the instants around the rebound defined as the origin of time, illustrated in the inset in (a), we notice that the cylinder is actually staying almost steady very near the bottom for almost $600\mu s$ after the contact, before starting to move away from it.
Some images every $200\mu s$ of the rebound are shown in Fig.\,\ref{fig:cavitation}(b), with symbols indicating the corresponding times in (a), a short movie of this dynamics is also provided in supplementary material as well (see supplementary material). One can observe the formation of a cavitation bubble at the location of contact, in between the cylinder and the bottom, as indicated by red arrows in 3 of the images. Its maximum horizontal extent is almost half the diameter of the cylinder. The nucleation of this bubble originates from very small bubbles at the surface of the cylinder and of the flat surface (visible in the image with the disc, for $t=0\mu s$). This is in good agreement with the crevice model for heterogeneous nucleation of bubbles in water \cite{atchley1989}, with nuclei trapped at the plastic surfaces being destabilized at rebound by pressure variations.

Similar observations have been obtained for all the rebounds with a cylinder of type C or D, in salt water and in uconn/water mixtures as well. The cavitation process seems to be related to the pressure drop generated at the rebound for sufficiently dense cylinders. The velocity at rebound is not the key element since no cavitation has been observed with cylinder of type B in water (falling faster than C in ucon/salt water).
Some tests on the influence of degassing the water of the tank  before the experiments have been done without noticing a difference in the process. Further investigations are needed, although they are out-of-scope for this manuscript.

To conclude, the cavitation bubble holds for about $200\mu s$ to $600\mu s$ (depending on the cases studied) and it is reproducible for similar conditions. It is found here that cavitation is a process that is also controlling the apparent time of contact $t_C$ for some cylinders (C and D), preventing it to be shorter. Altogether, it is thus found that the concept of contact time can be associated with several complex mechanisms.

\subsubsection{Non-normal rebound}

\begin{figure}
  \centering
  (a) \hspace{8cm} (b)\\
  \includegraphics[height=0.25\textheight]{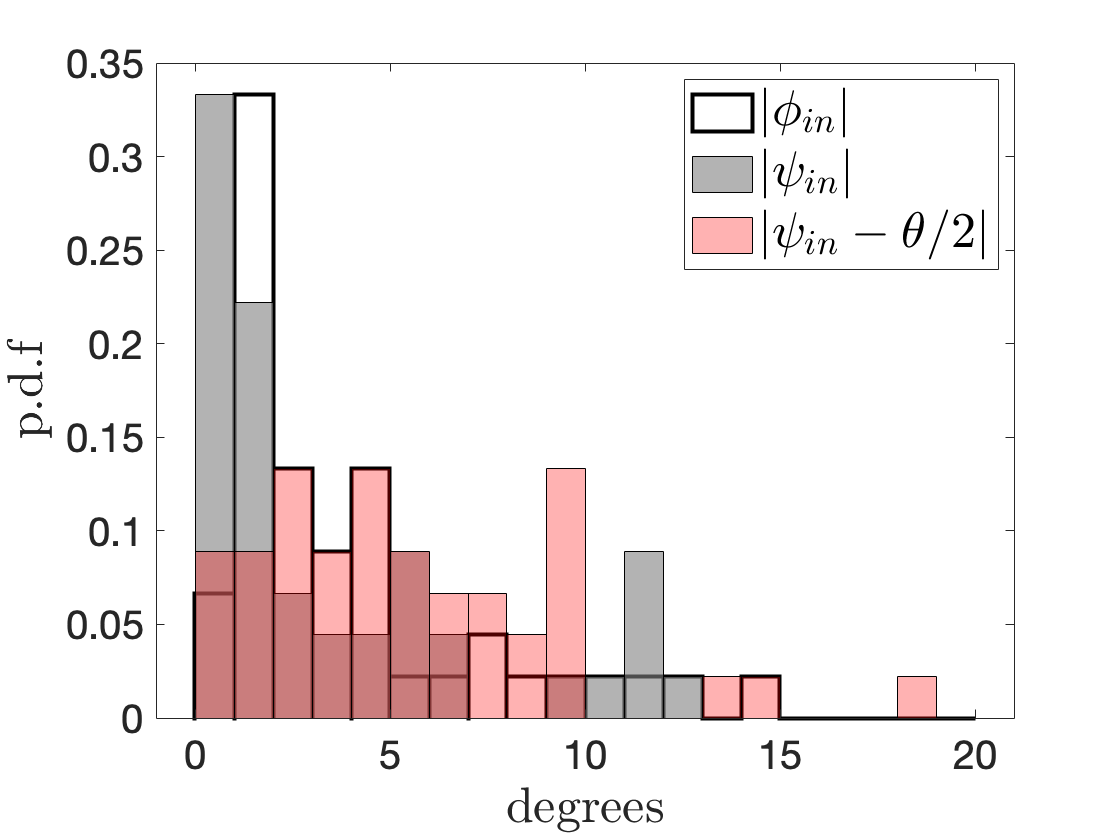} \includegraphics[height=0.25\textheight]{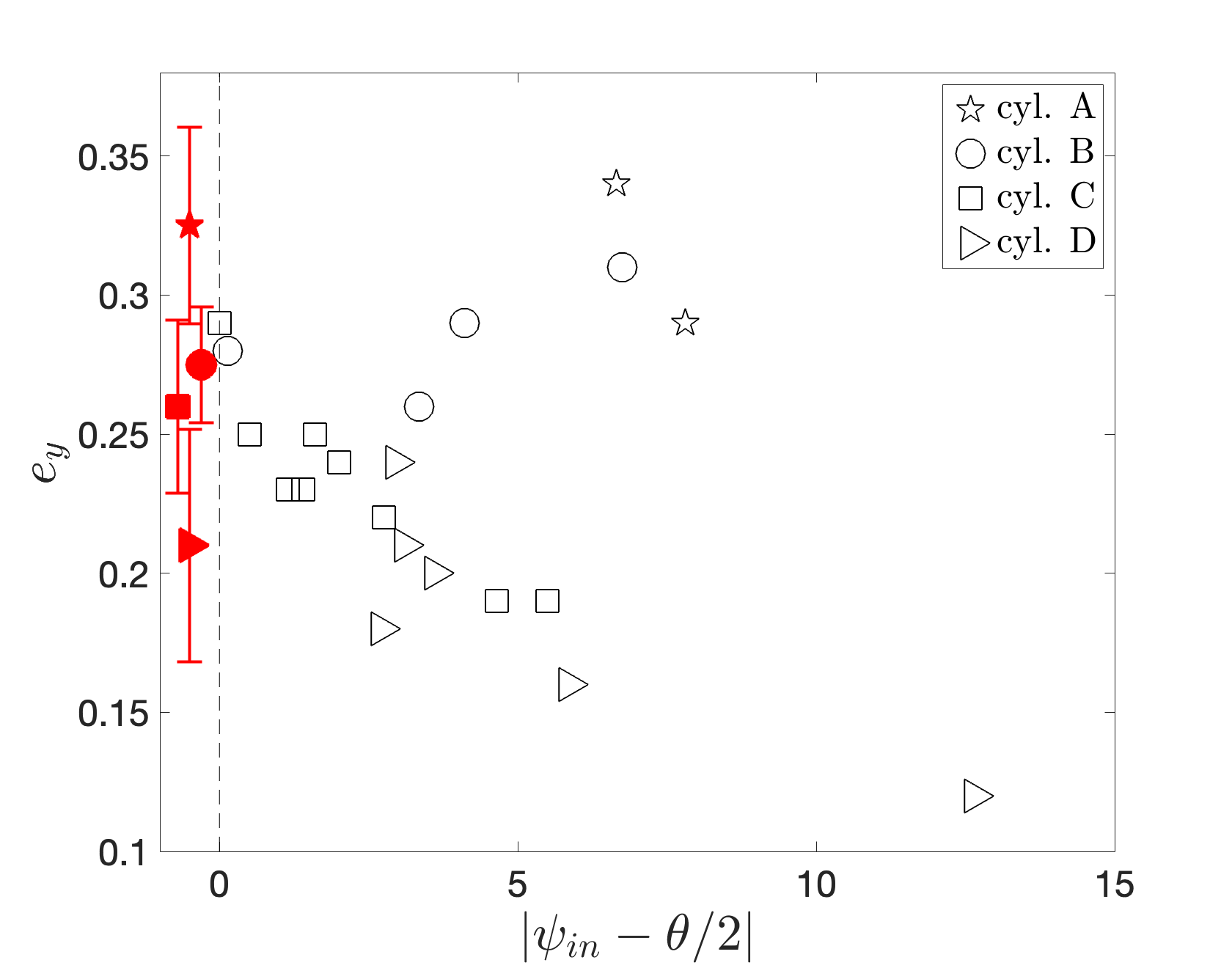}
\caption{(a) Distribution of the observed angles in air and in salt water at rebound of the cylinder itself $\phi_{in}$, the initial orientation $\psi_{in}$, and of the change in the orientation of its trajectory with respect to a perfect rebound with no friction $\psi_{in}-\theta/2$}. (b) coefficient of restitution $e_y$ in air, as a function of the rotation at rebound, sorted by the nature of the cylinder (symbols); the filled red symbols with error bars are the estimates of $e_{y_0}$ since $St>10^5$ for all cases.
\label{fig:St_e_exp_sorting}
\end{figure}
The other parameters that can influence the rebound are related to the orientation in the $(x-y)$ plane of the cylinder and its trajectory. 
For all experiments in air and salt water, we represent the distribution of the orientation of the object at contact ($\phi_{in}$), the orientation ($\psi_{in}$) of its trajectory, and the change in orientation with respect to a perfect rebound with no friction $\psi_{in}-\theta/2$ in Fig.\,\ref{fig:St_e_exp_sorting}(a). 
One can notice that the orientation at contact $\phi_{in}$ is peaked near $[0,1]$ degree indicating that in most cases the cylinder tail remains nearly vertical, although some non-vertical orientation can occur. The evolution of this orientation after rebound is almost unnoticeable for most cases, and there is no correlation of $\phi_{in}$ with the coefficient of restitution or other angles discussed below (not shown). 
The distribution of angle the trajectory of the cylinder before rebound is also peaked near $[0,1]$ degree. Nevertheless, some perturbations in the fluid tank or at release can induce some fluctuations in the fluid resulting in larger values of $\psi_{in}$. 
Although some mechanisms could relate the inclination of the object with the one of the trajectory (such as a lift force for instance), we did not find the values of $\psi_{in}$ and $\phi_{in}$ to be correlated.

Finally, the most important quantity to measure the non-normal aspect of the rebound is the distribution of the change in the orientation of the trajectory of the cylinder, $|\psi_{in}-\theta/2|$. The pdf shown is nearly flat for values between $0$ and $10$ degrees, with few cases larger than $10$ degrees. This quantity measures the importance of friction in the $x$-direction that can induce not trivial trajectories at rebound as discussed in \cite{Joseph2004}.

\subsubsection{Rebound in air}

In Fig.\,\ref{fig:St_e_exp_sorting}(b), we focus on the experiments done in air. They correspond to very large values of the Stokes number and are used to define a reference coefficient of restitution $e_{y_0}$ solely controlled by elasticity (red symbols). The estimates for $e_{y_0}$ are indicated with errorbars to account for the influence of the rotation of the trajectory at rebound. The values are slightly different for each object (decreasing from A to D), indicating that the inner core might play a part on the elastic response of each object. More specifically, the value of $e_y^0$ decreases with increasing mass of the object, which can also be explained by the dissipation increasing in the fixed plate at the bottom.

Finally, it is worth emphasizing that the order of magnitude for $e_{y_0}$ of $0.3$ is quite small for this regime with little dissipation in air. 
Although some complex solid-body dynamics could be considered to estimate the upper-bound for a rebound with translational/rotational energy transfer but no dissipation, here we consider this value to be mainly due to the energy dissipation in the solid bottom that was made of a large PVC block, rigidly maintained at the bottom of the tank with metal rods, as described in \ref{sec:exp_setup}. The unnoticeable vibrations of this structure are certainly the best explanation.

\subsubsection{Synthesis}

All the specificity of the apparent rebound of a cylinder discussed before can influence the coefficient of restitution in the vertical direction, and generate an important dispersion of the experimental observations. Nevertheless, we have not clearly identified a correlation between $e_y$ and the contact time, out-of-plane orientation or number of contacts at rebound, except for 3-contacts that always correspond to weaker values of $e_y$.
In the following, to compare with idealized numerical simulations, we will consider only observations for which $\psi_{in}-\theta/2$ strictly smaller than $6$ degrees, and with 1 or 2 contacts eventually. We could label these cases 'nearly-normal' rebounds of a cylinder.

\begin{figure}
\centering  \includegraphics[height=0.23\textheight]{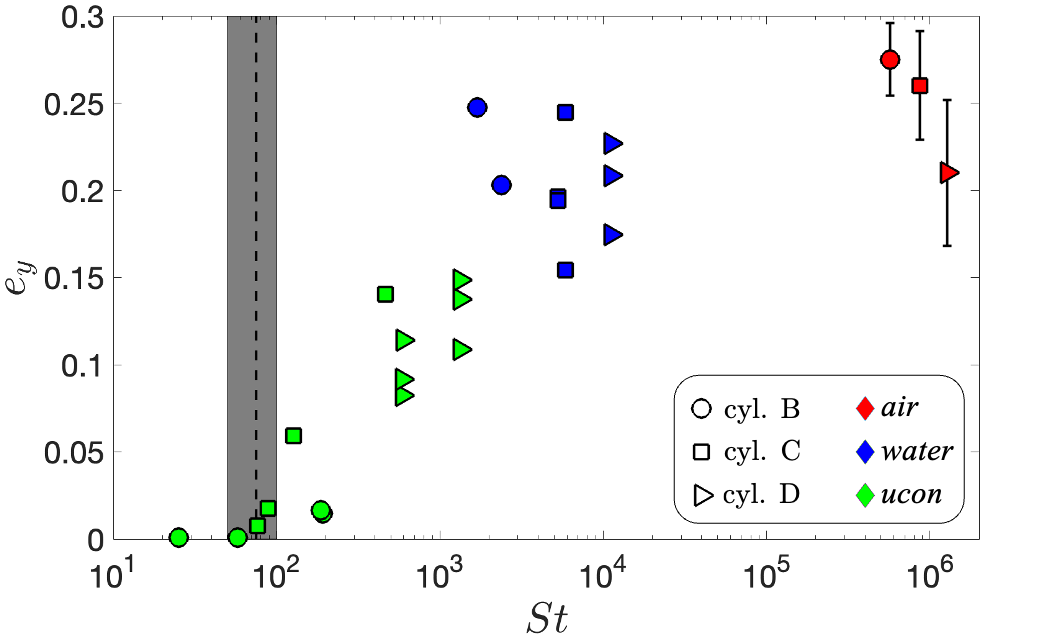}
\vspace{-0.25cm}
  \caption{The coefficient of restitution as a function of the Stokes number for experiments corresponding to nearly normal rebound. Colors of symbols indicate the fluid considered, while the shape refers to the properties of the cylinders. The critical value for rebound, $St_c = 75 \pm 25$, is indicated by the black dashed line with gray shaded area.}
\label{fig:ey_St_exp_alone}
\end{figure}

The coefficient of restitution $e_y$ obtained in these selected experiments, according to the above discussion, is shown as a function of $St$ in figure \ref{fig:ey_St_exp_alone}, the shape and color of symbols indicating the cylinder and fluid properties as described in tables \ref{tab:cylinders} and \ref{tab:fluids}. Error-bars in $e_y^0$ are discussed in the previous section, but we can also notice some dispersion of the results in salt water and in ucon/salt water. for which each symbol is a given experiment, a signature of the many (uncontrollable) parameters at play (cylinder inclination, pitching, cavitation,...).
The main result here is the estimation of a threshold $St_c$ above which rebound actually occurs. Above $St_c=75 \pm 25$, $e_y$ increases with $St$ until reaching $e_y^0$ for high enough values. Furthermore, it can be noticed that the "S-shape" that has been already evidenced in the past studies on spheres bouncing on a wall \cite{Joseph2001, Gondret2002, Davis1986} is still observed in the experiments on the falling cylinder. This suggests a possible applicability of the concept of previous established models, provided that those models are extended to take into account the specificity of the contact in the actual problem, compared to the contact occurring between a sphere and a plane. We will show in the following section, that it is sufficient to use the apparent (measured) contact time in order to predict the coefficient of restitution as a function of the Stokes number. \\

%%%%%%%%%%%%%%%%%%%%%%%%%%%%%%%%%%%%%%%%%%%%%%%%%%%%%%%%%%%%%%%%%%
\section{Numerical simulations of 2D effective bouncing}
\label{sec:Sim}

\begin{figure}
  \centering
  \setlength{\unitlength}{1mm}
\begin{picture}(180,160)
\put(0,50){\includegraphics[width=0.43\linewidth]{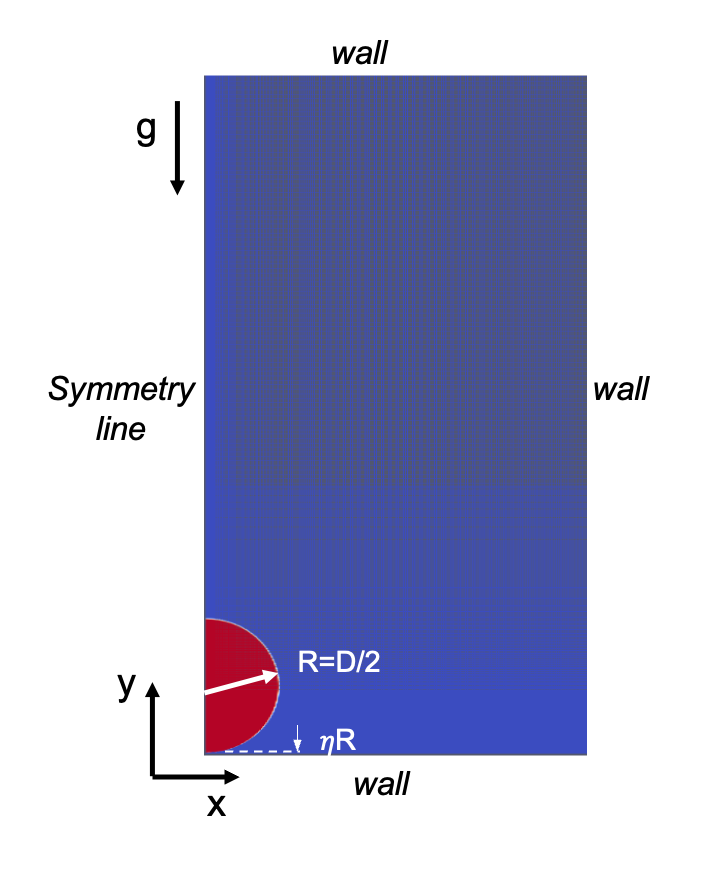}} 
 \put(75,90){\includegraphics[scale=0.58]{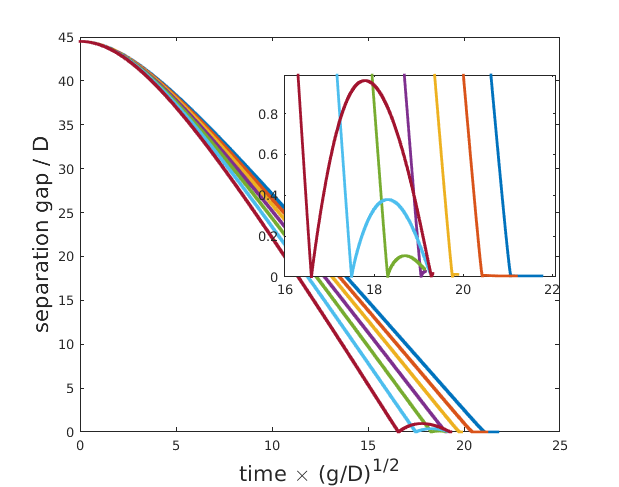}}
 \put(75,5){\includegraphics[scale=0.60]{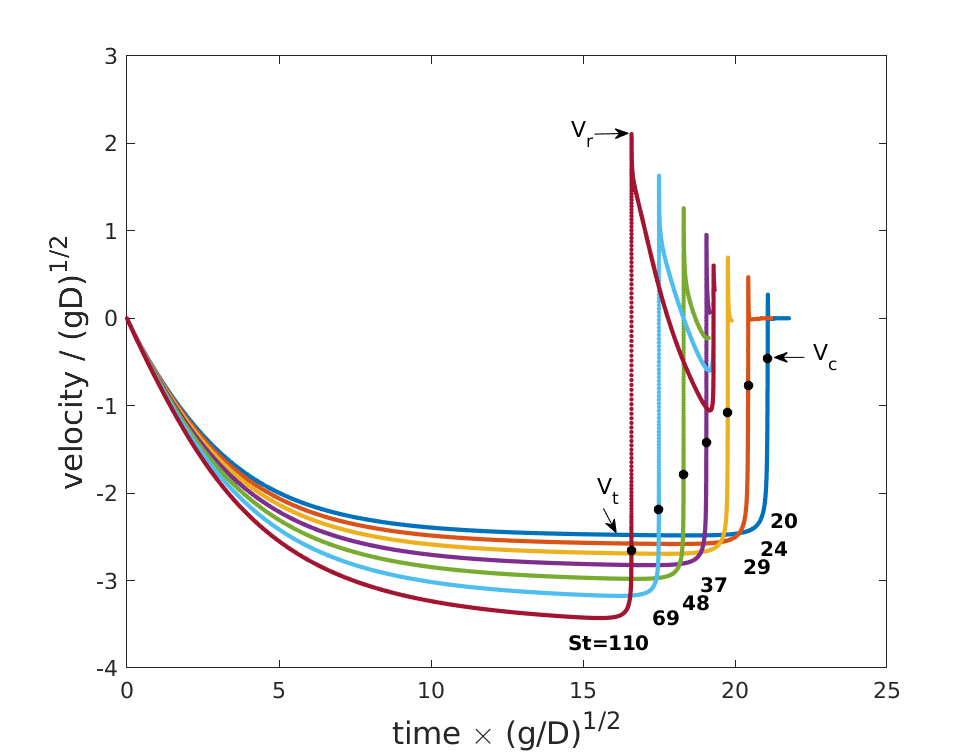}}
 \put(19,55){$0$}
 \put(16,138){$L_y$}
 \put(59,60){$L_x$}
 \put(5,138){(\textit{a})}
 \put(77,160){(\textit{b})}
 \put(77,74){(\textit{c})}
 \end{picture}
 \vspace{-1.25cm}
\caption{(\textit{a}) 2D numerical simulation setup, including the grid distribution and boundary conditions. Evolution in time of (\textit{b}) the separation gap between the cylinder surface and the wall and (\textit{c}) the cylinder velocity. The inset in (b) shows a zoom at small gaps. The colors correspond to different Stokes numbers indicated in (c). The black dots indicate the collision velocity $V_c$. The elasticity parameter used for those simulations is $\alpha=4$.}
\label{fig:num_setup}
\end{figure}
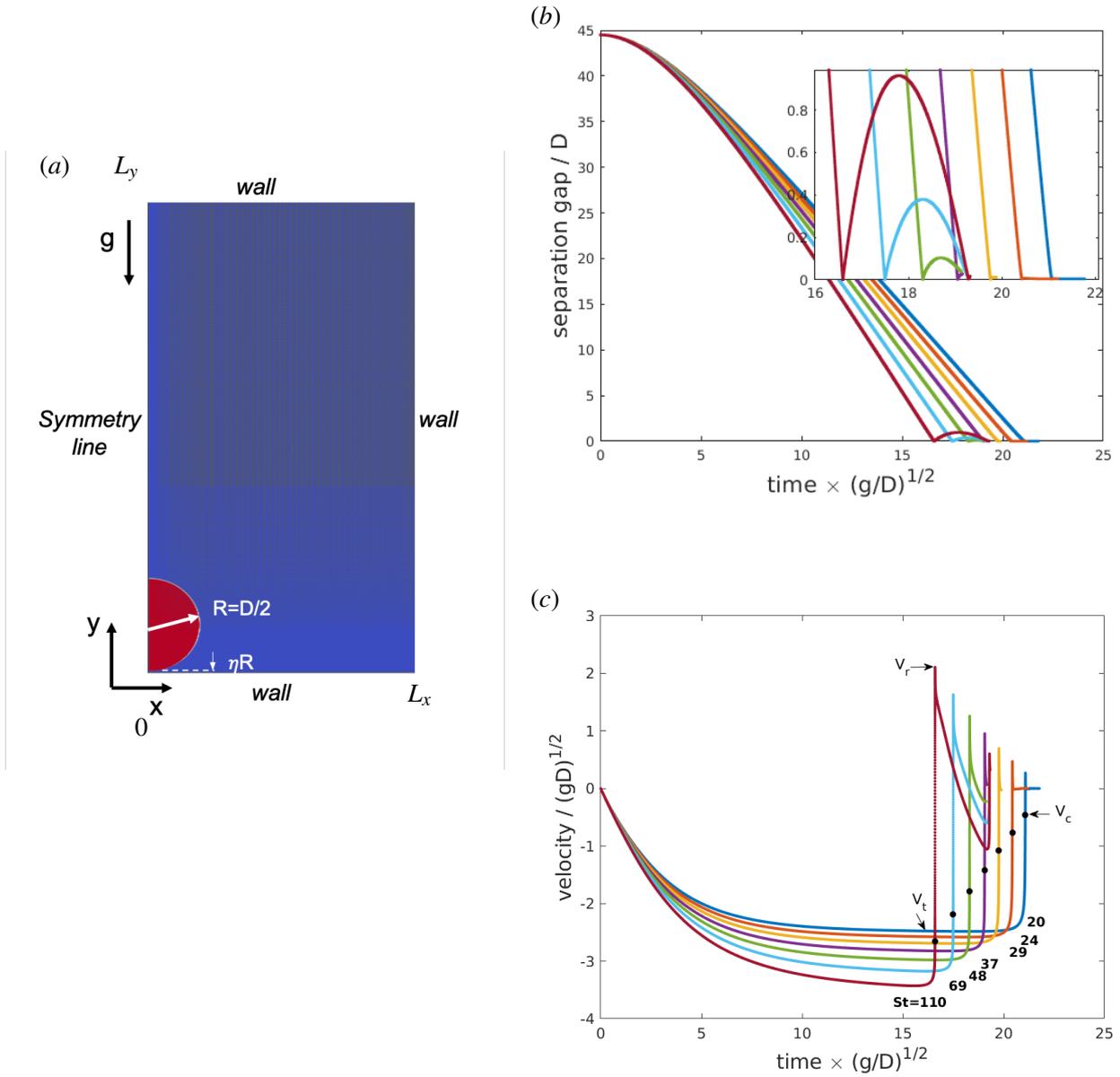

The numerical setup is sketched in figure \ref{fig:num_setup}(a). It consists of a two-dimensional closed vessel of dimension $(Lx,Ly)=(20R,100R)$ along the vertical $y$ direction and the horizontal $x$ direction, respectively, filled with a viscous fluid. No-slip velocity is imposed on three boundaries considered as solid walls. The fourth boundary $x=0$ (along the $y$ direction) is a line of symmetry (see figure \ref{fig:num_setup}(a) for details). A 2D cylinder of radius $R$ is freely settling under an imposed force $g$ along the negative $y$ direction. The cylinder center $(x_p,y_p) $initially placed at the line of symmetry is constrained to move along $y$ and thus remains at $x=0$. The position of the centre of the 2D cylinder is therefore denoted $(0,y_p)$, with $y_p$ depending on $t$. Note that using advantage of the properties of symmetry of this configuration, only half of the cylinder is represented and simulated here.

The coupling between fluid and 2D cylinder equations of motion is based on the Immersed Boundary Method, that we previously used to study the rebound on a wall of i) a sphere settling under a constant force \cite{izard2014modelling} and ii) a sphere carried by a wall-normal flow toward the stagnation point \cite{Li2020, Li2020b}. 
However in the present work, unlike our work on the settling sphere \cite{izard2014modelling}, we do not include any lubrication correction when the cylinder is close to the wall, as the motion of the fluid squeezed between the cylinder surface and the wall is resolved down to a scale relatively small compared with the cylinder radius $R$ as explained in \cite{Li2020, Li2020b}.  
In particular, near the contact region, $(x,y) \approx (0,0)$, a fine grid resolution is used: almost 20 grid points are employed in the region $y<0.01R$. This limits the minimal apparent roughness scale of the particle to a small fraction of the particle radius, as will be explained later on. Elsewhere, non-uniform Cartesian grid is used for computational efficiency. The size of the grid elements increases smoothly along positive $x$ and $y$ directions. At large distance from the wall, the spatial discretization ensures 30 grid points per cylinder diameter in the $y$ direction, and slightly more in the $x$ direction.

Simulations are carried out with constant cylinder diameter, and constant fluid and cylinder densities. Cylinder inertia is varied by changing the fluid viscosity and thus the Reynolds number while the density ratio is kept constant. The Reynolds and Stokes numbers defined from \ref{eq:Re} and \ref{eq:St} are varied in the range $[1-500]$. Note that using a symmetrical domain leads to a symmetric wake behind the cylinder. Even if wake instability shall be expected at the highest $Re$, this assumption remains acceptable as the settling time scale of interest for the bouncing is much smaller than the time scale required for the wake instability to take place, and therefore to affect the results presented here. 
During the settling stage, and after rebound, the time step is set to a small fraction of the settling time $dt\approx 0.001d/|V_t|$. 
During the collision stage, a smaller time step, equal to $1/40$ of the collision time estimated from eq.~\eqref{eq:Tc} as explained later.

The 2D cylinder settles from rest, its center being initially located at $y=90R$. It accelerates until the velocity reaches a terminal velocity $V_t$, balancing drag and apparent weight, prior to sudden deceleration due to a large hydrodynamic resistance close to the bottom wall (see the temporal evolution of the cylinder velocity in figure \ref{fig:num_setup}(b)). From the simulations, the cylinder terminal velocity $V_t$ is obtained from the temporal signal of the cylinder velocity (as shown in figure \ref{fig:num_setup}(b)). The plateau corresponding to the balance between drag and apparent weight is clearly observed for $St<100$, between the acceleration and deceleration stages. At larger $St$, the cylinder falling time is not sufficient to reach the plateau. In this case, the terminal velocity is then defined as the maximum (negative) velocity before the cylinder starts to decelerate. 

At the end of the deceleration stage the cylinder comes to rest at low inertia, i.e. small $St$. However at high inertia, larger $St$, the cylinder surface becomes critically close to the wall while the cylinder settling velocity remains finite. 
When the gap between the cylinder surface and the wall becomes smaller than a threshold $\eta R$, referred to as the apparent roughness length ($\eta$ is the contact roughness), solid contact is assumed to occur. The finite velocity of the particle at the contact onset called $V_c$ hereafter, will be taken as the particle velocity when $\delta_n = y_p - (1+\eta)R$ becomes negative. While the value of $\eta$ is set to few percents in the numerical simulations, we carefully verified that the fluid motion in the separation gap is fully resolved with more than 20 grid points during solid contact, in a way to capture correctly the lubrication effect without using sub-grid models. Solid collision is then accounted for in the numerical simulation by adding a contact force to the cylinder equation of motion (in the wall-normal direction), as soon as $\delta_n$ becomes negative. 
The contact force is modeled as a linear elastic force $F_{c,n} = -k_n \delta_n$ with spring stiffness $k_n$. The solid dissipation during the contact is neglected in order to maintain viscous lubrication as a unique source of energy dissipation. We use $\delta_n$ to represent the solid apparent overlapping during the contact which mimics elastic deformation. The spring stiffness per unit length is thus modeled as
\begin{equation}
\label{eq:kn}
    k_n = \frac{m\ \pi^2}{L \left( \alpha T_H \right)^2},
\end{equation}
where $m/L$ denotes the cylinder mass per unit length.
This relation between the spring stiffness, the object mass and the contact time results from the simple modeling of collision of objects with a harmonic oscillator assuming energy dissipation is negligible during the collision process.
With this definition, $\alpha T_H$ sets the characteristic time scale for the collision.  We were inspired by the Hertzian collision time scale used to interpret the experimental results for spherical particles colliding with a wall \cite{zenit1997collisional} (based on reference \cite{goldsmit1960theory}) and adapted $T_H$ to cylindrical particle is written as:
\begin{equation}
\label{eq:Tc}
    T_H = \left[ \left( \frac{m}{L} \ \frac{1}{E_{el}} \right)^2  \frac{R}{2 V_c} \right]^{0.2},
\end{equation}
where $E_{el}$ denotes the Young Modulus of the cylinder, assumed equal to that of the wall. Note that we could have considered constant values for $T_H$. This does not impact significantly the results and the main messages that we would like to draw in the following.

\begin{figure} 
\centering
\includegraphics[scale=0.7]{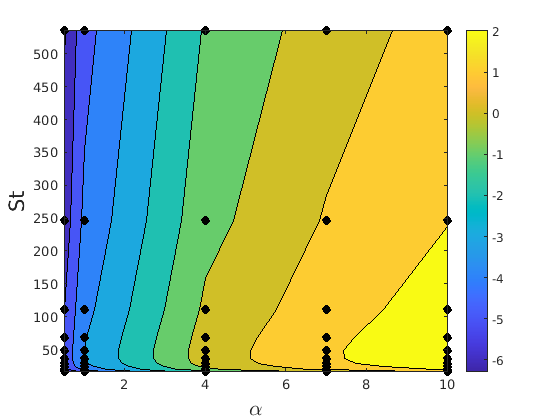}
\caption{Contour plot (in natural log-scale) of the softness dimensionless number $S=\frac{\mu V_c/(\eta R)}{E_{el}/\alpha^{2.5}}$ for all numerical simulations used for this study. The value of this parameter ranges between $10^{-3}$ (the darkest color) and $O(10)$ (the lightest color). The dots indicate the parameters used in the different simulations carried out with $\eta=0.01$. }
\label{fig:Capillary}
\end{figure}

If the collision between the cylinder and the wall was taking place in dry conditions (in the absence of viscous liquid), the contact duration would have been equal to $\alpha T_H$. However in viscous liquid, the contact duration is larger than this value , the difference being a decaying function with the Stokes number. The real collision time is measured \textit{a posteriori}, and we found that it is not very different from $\alpha T_H$, as it will be shown later. The parameter $\alpha$ allows to tune the contact elasticity (or in other terms the solid softness) in a practical way in the simulations. Thus the effective Young modulus is equal to $\frac{E_{el}}{\alpha^{2.5}}$. Softer (resp. stiffer) collisions occur for larger (resp. smaller) $\alpha$. A dimensionless number, $S=\frac{\mu V_c/(\eta R)}{E_{el}/\alpha^{2.5}}$, is built in a way to compare the shear stress of the flow in the separation gap during the collision process (estimated by $\mu V_c/(\eta R)$ and varies with the Stokes number) and the solid elasticity. The meaning of this dimensionless number is similar to that defined in the elastohydrodynamic theory of \cite{Davis1986}. Figure \ref{fig:Capillary} shows the contour plot of $S$ corresponding to the conditions where numerical simulations have been carried. The range of $S$ obtained here is close by construction to the capillary number corresponding to slightly deformable drops. Figure \ref{fig:Capillary} suggests that $S$ depends weakly on the Stokes number at small $\alpha$, which means that the contact softness is mainly tuned by the elasticity parameter $\alpha$.

From the simulations we extract the terminal velocity $V_t$, the velocity $V_c$ at the onset of collision process (at the instant where $\delta_n$ becomes negative) and the velocity $V_r$ at the end of the collision process (at the instant when $\delta_n$ becomes positive again). In the simulations, the "dry" coefficient of restitution is equal to one (the contact force $F_{c,n}$ does not contain any damping term). The effective coefficient of restitution is defined as the ratio $e=-V_r/V_t$ which can be written as the product of two contributions, prior to solid contact $V_c/V_t$ and during contact $-V_r/V_c$. 
These two contributions are plotted as a function of $St$ in figure \ref{fig:Vc_Vt-Vr_Vc}. Panel a) in this figure shows that  $V_c/V_t$ increases with $St$, which suggests that, at the onset of rebound, the cylinder kinetic energy increases with $St$. Panel b) shows that $-V_r/V_c$ is close to 1 and almost independent of $St$ when the elasticity parameter is small. The magnitude of $-V_r/V_c$ is significantly decreased when the elasticity parameter is increased (softer contact). Figure \ref{fig:St_e_num_model} shows the separate effect of the cut-off length $\eta R$ (panel a) and elasticity parameter $\alpha$ (panel b) on the coefficient of restitution, when the Stokes number is varied. This figure allows showing that the coefficient of restitution depends on both parameters that are linked to two different physical processes: the elasticity parameter $\alpha$ which mainly controls the effective contact time of the collision and therefore the contact softness/rigidity, and the cut-off length $\eta$ that sets the collision onset, akin an apparent roughness. The larger the apparent roughness, the earlier the no-rebound to rebound transition takes place (the cut-off length $\eta$ mostly affects the critical Stokes $St_c$). The larger the elasticity parameter, the longer is the contact time, and the smaller is the restitution coefficient for a given Stokes number.

\section{Model for the restitution coefficient}
\label{sec:model}

This section explains the basis of a simple model constructed to allow the prediction of both the cylinder velocity at the onset of collision $V_c$, and the rebound velocity $V_r$ when the solid contact is completed. This model follows closely the one outlined in the work of Izard et al. \cite{izard2014modelling} (considering the rebound of a sphere on a wall). 

\begin{figure} 
\centering
\setlength{\unitlength}{1mm}
\begin{picture}(170,65)
\put(0,0){\includegraphics[width=.49\textwidth]{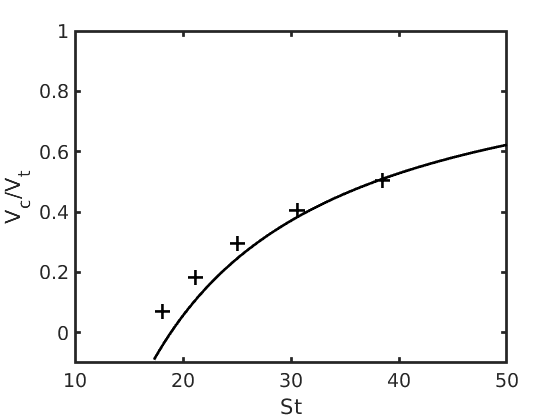}}
\put(90,0){\includegraphics[width=.49\textwidth]{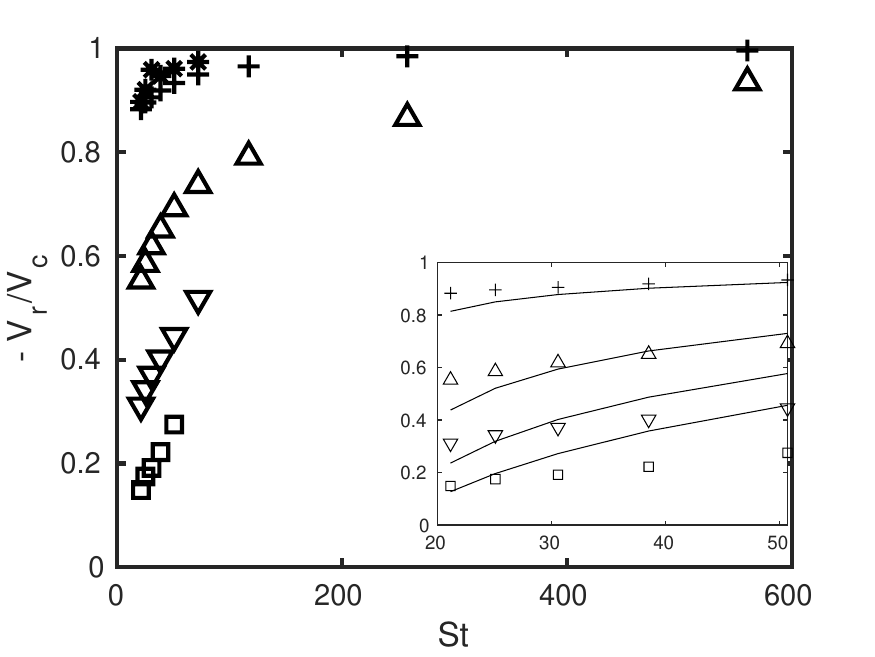}}
\put(0, 60){(\textit{a})}
\put(90, 60){(\textit{b})}
\end{picture}
\caption{ (a) Collision-to-terminal velocity ratio $V_c/V_t$ and (b) rebound-to-collision velocity ratio $-V_r/V_c$ as a function of the Stokes number. $V_c$ is the particle velocity at the collision onset, for $\eta=0.01$. The symbols correspond to numerical simulations. The solid line in panel a) predicts $\frac{V_c}{V_t}$ from eq. \ref{eq:Vc_Vt}. In panel (b) different symbols correspond to different $\alpha$: stars, plus, up triangles, down triangles and squares correspond to $\alpha=0.5$, 1, 4, 7 and 10, respectively  (from rigid to soft cylinder). The solid lines in the inset correspond to $-\frac{V_r}{V_c}$ obtained from eq. \ref{eq:Vr_Vc0} using a collision time $\tau = \pi \sqrt{\frac{m^*}{k_n}}$ and $\alpha=$ 1, 4, 7 and 10 (from top to bottom).}
\label{fig:Vc_Vt-Vr_Vc}
\end{figure}

First, let us attempt to estimate the ratio between the collision velocity $V_c$ and terminal velocity $V_t$, by writing the equation of motion of the cylinder during the deceleration stage. While unsteady forces that the cylinder might experience during the deceleration stage are not readily available in the regime of interest (Stokes number ranging between 10 and 100), we write a simplified equation of motion of the cylinder, in a way to predict the motion at the leading order. For this, we assume that the hydrodynamic force experienced by the particle is the superposition of the drag experienced by a cylinder moving in an unbounded fluid, the viscous lubrication associated with the pressure divergence as the gap between the cylinder surface and the wall $ \zeta=(y_p-R) \rightarrow 0$ and an added mass contribution

\begin{equation}
\label{eq:motion}
m^* \frac{dV}{dt} = F_{lub}.
\end{equation}

In eq.~\eqref{eq:motion}, $V$ denotes the cylinder velocity and $m^*=\rho^* \pi R^2$ corresponds to the cylinder mass per unit length accounting for the added mass effect. $\rho^*=\rho_p\left(1+C_M\rho_f/\rho_p\right)$ corresponds to the cylinder apparent density accounting for the added mass effect ($C_M$ refers to the added mass coefficient). Here, the gravity force and steady drag corresponding to a cylindrical body moving in an unbounded fluid are omitted, assuming that their balance is independent of the cylinder position. In other words, eq.~\eqref{eq:motion} describes the departure from the equilibrium (where the settling velocity equals the terminal velocity $V_t$) due to the presence of the wall. The lubrication force (per unit length) experienced by a 2D cylinder moving toward a wall with a given velocity and assuming the flow motion is quasi-steady follows (see appendix \ref{sec:app_Flub} and \cite{jeffrey1981slow}):
\begin{equation}
\label{eq:Flub}
F_{lub} = - \frac{6}{\sqrt{2}} \pi \mu \left(\frac{R}{\zeta}\right)^{3/2} \frac{d\zeta}{dt},
\end{equation}
where $d\zeta/dt$ corresponds to the particle instantaneous velocity in a Eulerian frame of reference, i.e. $d\zeta/dt=V$. Although this expression is strictly valid at $\zeta<<R$, we assume that it applies continuously from a distance $\zeta=O(10R)$ (where $F_{lub}$ tends to zero) until very small $\zeta$. Of course in the region $\zeta=O(R)$ where the particle decelerates, this approximation is not valid. An accurate force balance on a cylindrical particle approaching a wall, where unsteady contributions are accounted for, would be necessary, however not available currently. \\ 

Integration of \eqref{eq:motion}-\eqref{eq:Flub} from a distance where the cylinder starts to decelerate ($V\approx V_t$ at $\zeta\approx R$) to the wall where collision occurs ($V=V_c$ at $\zeta= \eta R$), assuming $\eta \ll 1$ leads to (see appendix \ref{annexe:vc_vt} and in a similar way the work of Davis et al. \cite{Davis1986}):

\begin{equation}
\label{eq:Vc_Vt}
\frac{V_c}{V_t} = \left[ 1 - \frac{St_c}{St} \right], \quad St_c = \frac{8}{3\sqrt{2}}\eta^{-1/2},
\end{equation}

\noindent $St_c$ being a critical Stokes number for bouncing, i.e. over which $ V_c \ne 0$. Then, when $St<St_c$, the cylinder energy is entirely damped away before it reaches the wall, i.e. $V_c/V_t \rightarrow 0$ when $\zeta\rightarrow \eta R$. However above the rebound onset, i.e. $St>St_c$, the ratio $V_c/V_t$ increases with particle inertia. Its value extracted from the numerical simulations is compared with the solution of eq.~\eqref{eq:Vc_Vt} for varying $St>St_c$ in figure \ref{fig:Vc_Vt-Vr_Vc}(a) (symbols correspond to numerical simulation and solution of~\eqref{eq:Vc_Vt} is plotted as solid line). Despite its obvious  simplicity to obtain a theoretical solution, the estimation of $\frac{V_c}{V_t}$ by the model~\eqref{eq:Vc_Vt} is acceptable.

\begin{figure}
  \centering
\setlength{\unitlength}{1mm}
\begin{picture}(180,65)
\put(0,0){\includegraphics[width=0.49\textwidth]{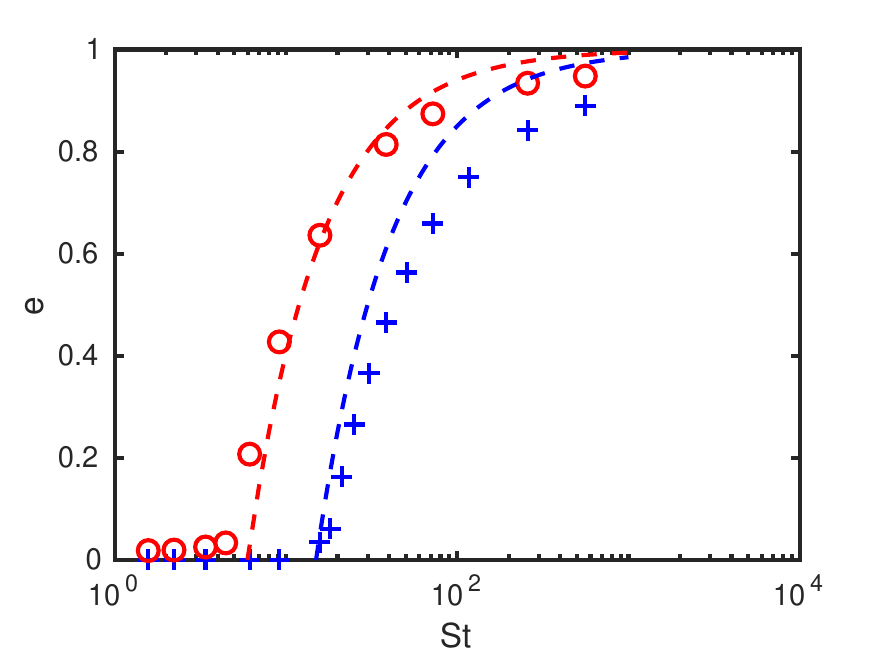}}
\put(90,0){\includegraphics[width=0.49\textwidth]{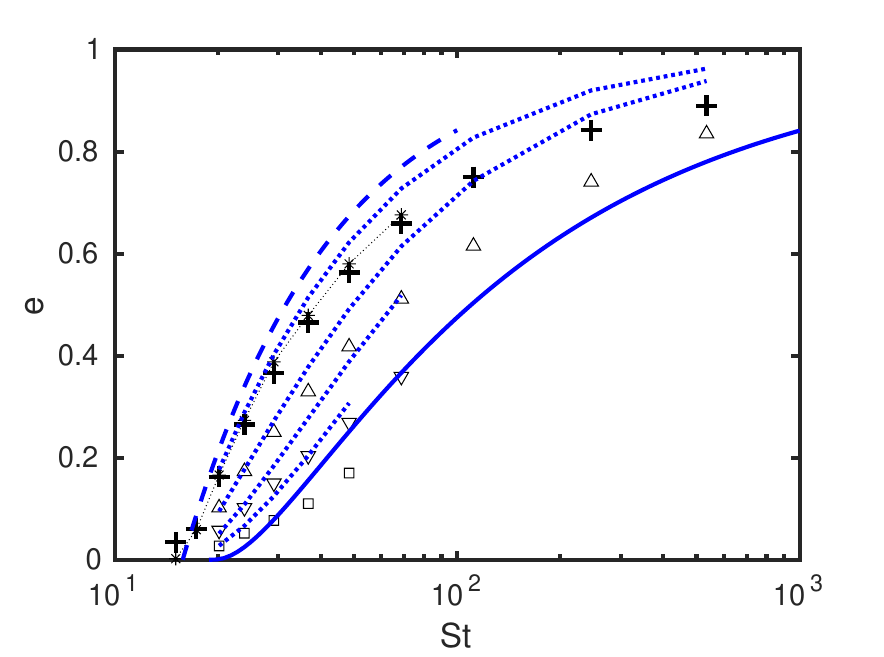}}
\put(14, 54){(\textit{a})}
\put(104, 54){(\textit{b})}
\end{picture}
  \caption{Coefficient of restitution as a function of the Stokes number; the symbols are from numerical simulations and the dashed lines correspond to $e_{RB}$ (eq. \ref{eq:e_RB}) assuming that the cylinder is infinitely rigid and that the near-wall hydrodynamic force during settling is dominated by lubrication. Panel (a) shows evidence of the role of apparent surface roughness: the blue plus and empty red circles correspond to $\eta=0.01$ and $0.1$, respectively, obtained with the elasticity parameter $\alpha=1$. Panel (b) evidences the role of elasticity: the stars, plus, up triangles, down triangles and squares correspond to simulations carried out with $\alpha=0.5$, 1, 4, 7 and 10, respectively and apparent roughness $\eta=0.01$. The dashed line is reported from panel (a). The solid line corresponds to $e_{Iz}$ (eq. \ref{eq:e_Iz}) assuming a balance between elastic deformation and lubrication forces during contact. The dotted lines are calculated from $e=-\frac{V_c}{V_t}\times\frac{V_r}{V_c}$, where $\frac{V_r}{V_c}$ is obtained from eq. \ref{eq:Vr_Vc0} with $\tau = \pi \sqrt{\frac{m^*}{k_n}}$ like in figure \ref{fig:Vc_Vt-Vr_Vc}b. These dotted lines correspond to $\alpha=$ 1, 4, 7 and 10, from top to bottom (from rigid to soft contact). }
\label{fig:St_e_num_model}
\end{figure}

Above the rebound onset, the cylinder velocity at the end of the collision $V_r$ is obtained from the equation of motion of the cylinder, accounting for viscous lubrication and contact force. For the latter, we consider the elastic force $F_{c,n}$ identical to the one used in the numerical simulations. Again, dissipation in the solid is neglected. This leads to the damped oscillator equation, written in terms of the overlapping distance with respect to the contact cut-off length scale $\delta_n= \zeta - \eta R$:
\begin{equation}
m^* \frac{d^2\delta_n}{dt^2} + \lambda \frac{d\delta_n}{dt} + k_n\delta_n = 0 \quad \mbox{with} \quad \lambda = \frac{6}{\sqrt{2}} \pi \mu \left(\frac{1}{\eta}\right)^{3/2}.
\label{eq:damp_oscill}
\end{equation}
The rebound velocity $V_r$ can then be estimated as $V_r = d\delta_n/dt$ at $t=\tau$, where $\tau$ denotes the time at which the overlapping distance is again equal to the cut-off length $\eta R$. Based on this definition, $\tau$ represents the solid contact time. The obtained rebound-to-collision velocity ratio reads
\begin{equation}
\label{eq:Vr_Vc0}
\frac{V_r}{V_c} = - exp \left[ - \frac{\lambda \tau}{2m^*}\right].
\end{equation}
According to this model, the ratio $V_r/V_c$ depends on the particle stiffness via the solid contact time $\tau$. The longer is the contact time, the smaller is the magnitude of the rebound velocity. 
Except very close to the rebound threshold, where viscous effects are important, the contact time is controlled by contact elasticity and it is reasonable to approximate it by $\tau = \pi \sqrt{m^*/k_n}$. 
Figure \ref{fig:Vc_Vt-Vr_Vc}(b) shows the ratio $V_r/V_c$ obtained from both numerical simulations and eq. \eqref{eq:Vr_Vc0}, for different elasticity parameter $\alpha$, ranging from $0.5$ to $10$. 
Overall, the model captures the trend found in the simulations, i.e. the ratio $V_r/V_c$ increases with $St$ and decreases with the elasticity parameter. The overestimation of  $V_r/V_c$ by the model is likely due to inertial effects that are not accounted for in eq.~\eqref{eq:damp_oscill}, as for instance the contribution of the wake behind the cylinder to the dynamics during motion reversal. 

Finally, the effective coefficient of restitution is defined as the negative product of both velocity ratios
\begin{equation}
   e=-\frac{V_c}{V_t}\times\frac{V_r}{V_c} = \left[ 1 - \frac{St_c}{St} \right] exp \left[ - \frac{\lambda \tau}{2m^*}\right]
   \label{eq:e_general}
\end{equation}
With this way of writing the restitution coefficient, we account for both the cut-off parameter (through $St_c$) and the contact elasticity (through the contact time $\tau$). We are therefore left with the description of the link between the contact time and the physical processes that come into play during bouncing. From figures \ref{fig:St_expe_model}b and d that will be discussed more deeply below, it can be suggested that this contact time is mainly dependent on the elasticity parameter $\alpha$ in the numerical simulations, whereas it depends on many complex phenomena like cavitation and cylinder pitching in the experiments, with a possible rationalization as a function of $St/St_c$.

In figure \ref{fig:St_e_num_model}b, the coefficient of restitution obtained from numerical simulations is compared with eq. \ref{eq:e_general} (displayed in dotted lines), in which the contact time was obtained posteriori in each simulation (the time during which the contact force is turned on). The agreement between numerical results and eq. \ref{eq:e_general} is relatively good. Indeed an increase of $\alpha$, i.e. softer contact, leads to longer contact duration and thus to a more significant contribution of the viscous dissipation to the overall energy loss during the collision process ($\frac{V_r}{V_c}$ decreases). Consequently the slope of the $e-St$ curve is lowered, and the transition of the coefficient of restitution from $0$ to $1$ is expected to take place on a much wider range of Stokes numbers. This relatively good agreement suggests that the proposed simple model contains the required ingredients to capture the energy loss during the 2D cylinder bouncing. 

In the next section we will consider several possibilities of modeling the solid contact time in order to obtain explicit expressions for the coefficient of restitution, depending on the governing physical process. Before that, let's consider the limit where that the contact time is infinitely small, i.e. $\tau\rightarrow 0$, then there is no energy loss during the collision, ($-V_r/V_c \rightarrow 1$). In that case, the coefficient of restitution called $e_{RB}$ (RB standing for rigid body) becomes equal to:
\begin{equation}
    e_{RB} = \left( 1-\frac{St_c}{St} \right).
    \label{eq:e_RB}
\end{equation}

\noindent $e_{RB}$ is displayed with dashed lines in figure \ref{fig:St_e_num_model}(a), with $\eta=0.01$ (blue) and $\eta = 0.1$ (red). One clearly observes that the $e-St$ curves obtained from numerical simulations with the most rigid cylinders ($\alpha=0.5$ and $\alpha=1$) follow closely the rigid body model, as long as the appropriate $St_c$, set by the apparent contact roughness modeled by $\eta$, is used.

%%%%%%%%%%%%%%%%%%%%%%%%%%%%%%%%%%%%%%%%%%%%%%%%%%%%%%%%%%
\section{Discussion on model parameterization }\label{sec:discussion}

\begin{figure}
  \centering
  (a) \hspace{8cm} (b)\\  
  \vspace{-0.0cm}
\includegraphics[trim=5 0 10 50,clip,height=0.24\textheight]{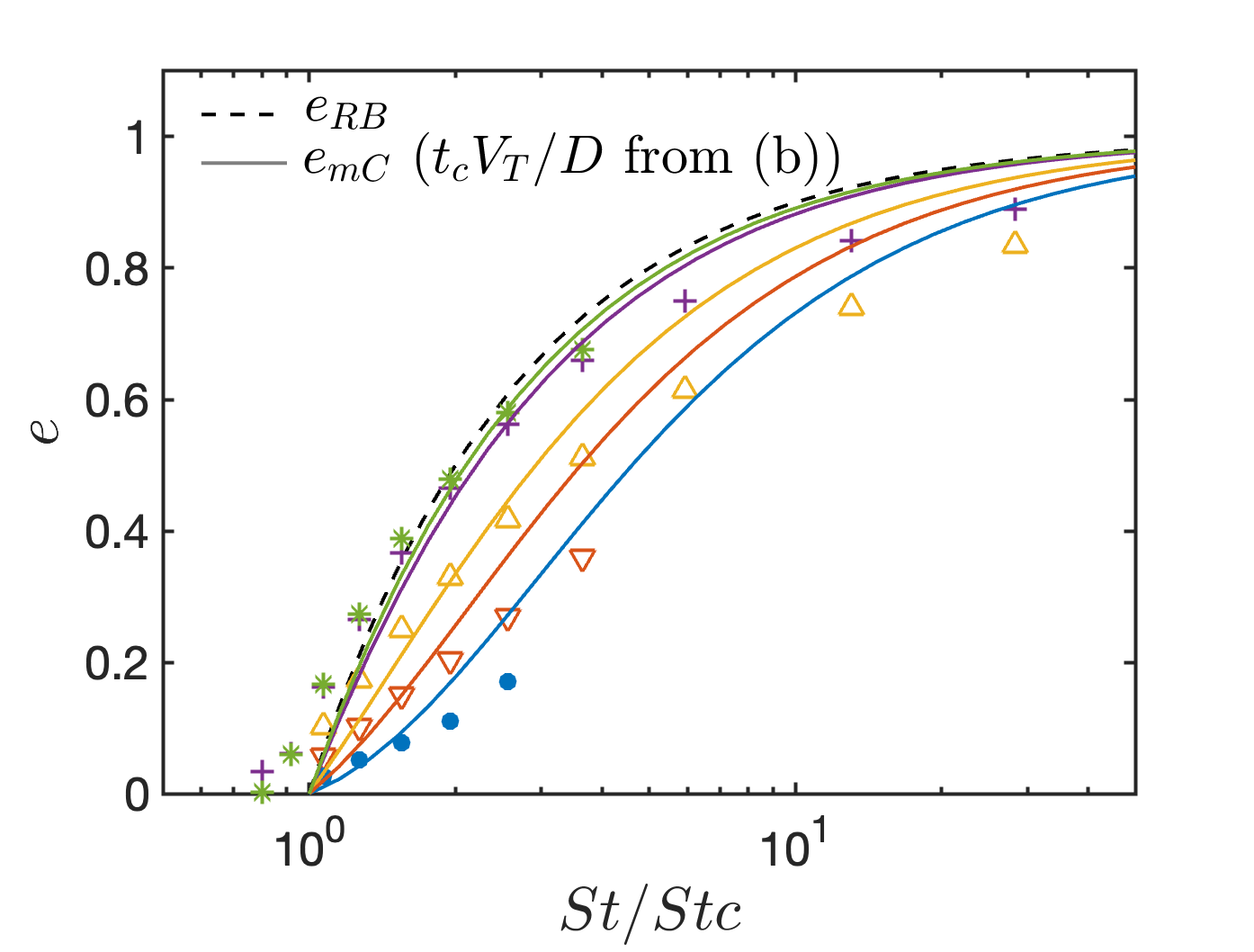}
  \includegraphics[trim=5 0 10 50,clip,height=0.24 \textheight]{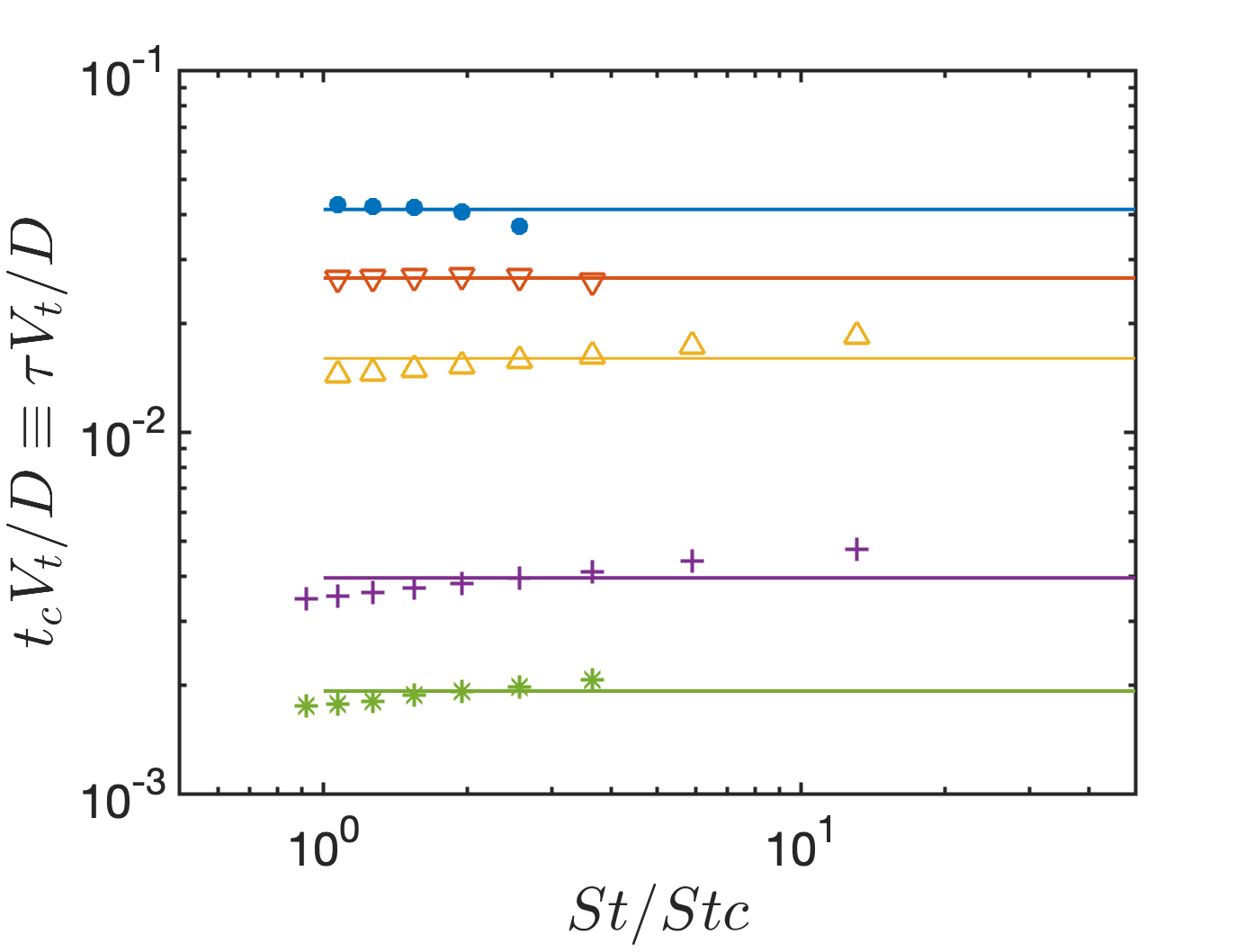}\\
  (c) \hspace{8cm} (d)\\  
  \vspace{-0.0cm}
  \includegraphics[trim=5 0 10 50,clip,height=0.24 \textheight]{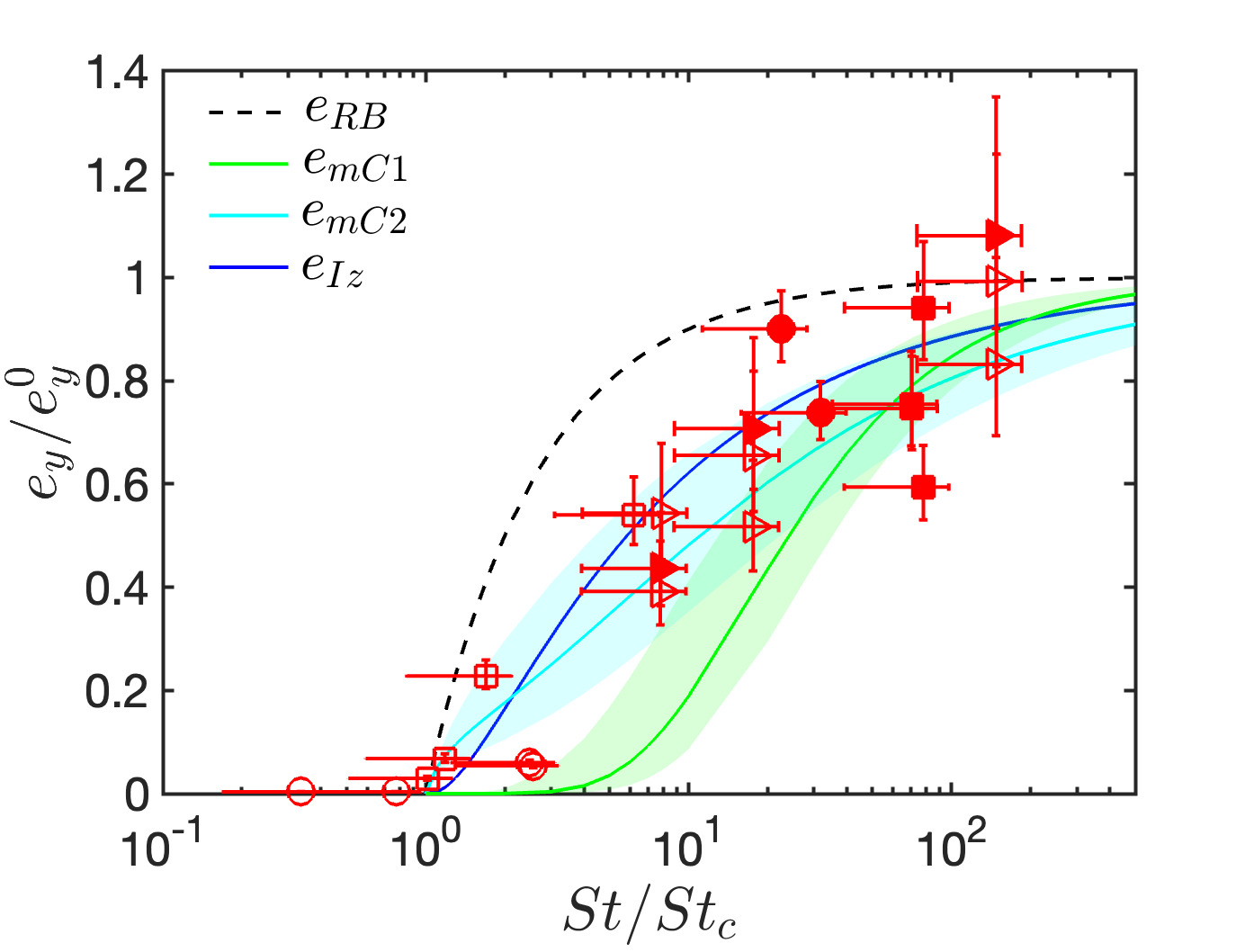}
  \includegraphics[trim=5 0 10 50,clip,height=0.24 \textheight]{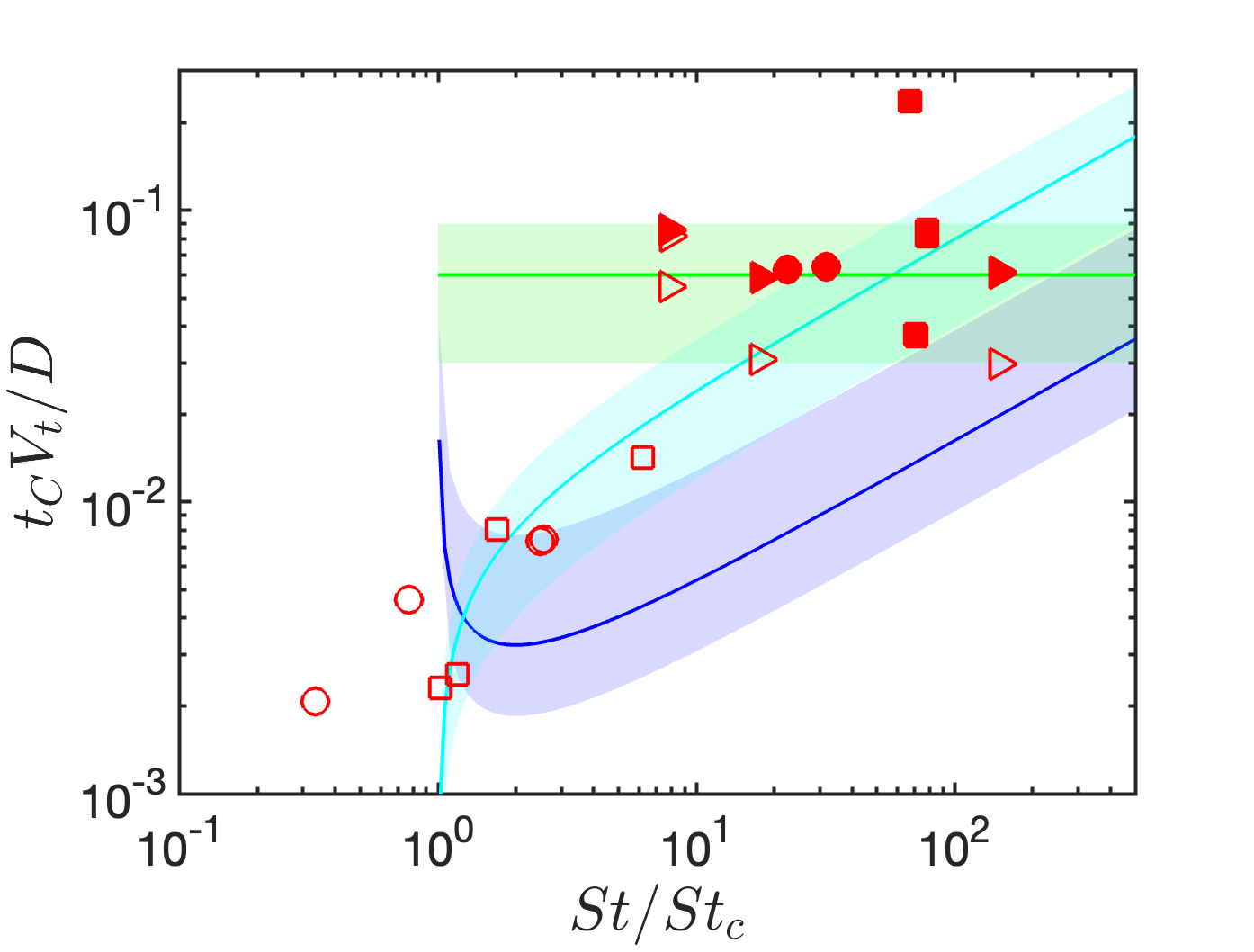}

  \caption{
  (a) \& (c) The coefficient of restitution normalized by its value in air as a function of the Stokes number and (b) \& (d) the non-dimensional contact time as a function of the Stokes number scaled by $St_c$. Symbols in (a) \& (b) are obtained from numerical simulations, and similar to figure \ref{fig:St_e_num_model}(b), obtained with different elasticity parameters $\alpha$. Red symbols in (c) \& (d) correspond to experiments as in figure \ref{fig:ey_St_exp_alone}. Vertical errorbars for experiments in (c) are associated to uncertainties in $e_y^0$ and horizontal ones to uncertainties in $St_c$ (not reproduced in panel d). Empty (resp. filled) symbols are for 1-contact (resp. 2-contact) rebounds. The black dashed line in (a) \& (c) is for $e_{RB}$. In (a) colored lines correspond to model \eqref{eq:modelrestitution} with constant dimensionless contact time $t_cV_t/D$ obtained from (b) (lines, same color legend). In (c) \& (d), the blue solid line is for $e_{Iz}$ and $\tau_{Iz}$ (eqs. \ref{eq:e_Iz} and \ref{annexe:tc_eIz}) with $\eta \in [3.6 - 15]\,10^{-4}$, while the green and cyan lines correspond to models $e_{mC1}$ and $e_{mC2}$ (eqs. \ref{eq:e_model1} and \ref{eq:e_model2}) with $St_c=75$ ($\gamma=2$) and apparent non-dimensional contact time being constant at $(6\pm3)\times10^{-2}$ or fitted by $(8\pm 4)\times10^{-3}\sqrt{St/St_c-1}$ respectively. Shaded areas around the models describe the uncertainties on the fitting parameters or the value of $St_c$.}
\label{fig:St_expe_model}
\end{figure}

%\begin{figure}
%  \centering
%%  \includegraphics[width=0.65\textwidth]{ey_ov_ey0_St_exp_models_label_v3.png}
%  \includegraphics[height=0.23\textheight]{Tc_e_with_exp.png}
%  \caption{Diagram showing the coefficient of restitution versus the contact time from both simulations and experiments. The symbols are identical to figure \ref{fig:St_expe_model}. } 
%\label{fig:Tc_e}
%\end{figure}
The study described in the previous section with a 2D cylinder suggests that (i) the increase of dissipation associated with the contact elasticity is well captured by a model based on solid contact time scale $\tau$, leading to relation~\eqref{eq:Vr_Vc0} for the velocity ratio $V_r/V_c$, and that (ii) the cut-off length $\eta$ impacts mainly the velocity ratio $V_c/V_t$ following relation~\eqref{eq:Vc_Vt}. We will examine here to which extent these concepts apply to our experimental measurements. The panels in figure \ref{fig:St_expe_model} display the variation of the coefficient of restitution (a) \& (c) and the contact time $t_c$ (b) \& (d) as a function of the Stokes number, from numerical simulations and experiments respectively. Note that for numerical simulations the contact time $t_c$ is the one discussed in the model \eqref{eq:Vr_Vc0}, i.e. $t_c\equiv \tau$. The Stokes number is scaled by the critical Stokes number corresponding to the onset of bouncing in different cases.
Regarding the results from experiments, we remind the reader that only those obtained from observations when the rebound is nearly-normal, and when the contact dynamics is with 1 or 2 contacts, are kept in this figure \ref{fig:St_expe_model}. This induces some dispersion of the results, due to the various processes at contact between the falling object and the wall, as classified in section \ref{sec:exp_results}. This is different from measurement uncertainties, with vertical error-bars associated with uncertainties in $e_y^0$ and horizontal ones with uncertainties in $St_c$, when changing its value from $50$ to $100$.

Let's examine in figure \ref{fig:St_expe_model}(c) the coefficients of restitution from experiments. Despite the dispersion in experiments, we can derive a general trend based on previous modelling. The departure of $e$ and $e_y/e_y^0$ from zero near $St_c$ is assumed to be controlled by the apparent contact roughness, imposed in the numerical simulations and associated with the surface properties of the material considered in the experiments (PVC in our case). If single rigid contact is first considered, the coefficient of restitution could be estimated from the rigid body approximation, i.e. following eq.\,\eqref{eq:Vc_Vt}, shown by the black dashed line.
This model is only characterized by the threshold in Stokes number. According to eq.\,\eqref{eq:Vc_Vt}, the numerical value of $St_c\cong 18.9$ is imposed by the value of $\eta=0.01$ used here, while the experimental estimate of $St_c=75$ would correspond to a cut-off length of $\eta = 6.3\times 10^{-4}$.
% Comment here on the validity of Stc formula
It is worth noting here that this expression for $St_c$ (obtained with the definition of the Stokes number in eq.\,\eqref{eq:St} inspired from the falling sphere) should be adjusted to account for the shape of the experimental object, i.e. a cylinder with tail. However, the form of the relationship between $\eta$ and $St_c$ still holds as 
\begin{equation}
\label{eq:Stc}
St_c = \gamma\eta^{-1/2},
\end{equation}
where $\gamma$ shall be a shape dependent constant of order unity ($\gamma \simeq 1.88 $ for the cylinder without tail, as in simulations).
Whatever the value of this constant, the value of $\eta$ does not seem to reflect the range of $\delta z/R$ reported for experiments in figure \ref{fig:timelapse}(c). Varying $St_c$ from $50$ to $100$ (with $\gamma=2$) would lead to $\eta\in [4.0\times10^{-4} - 1.6\times10^{-3}]$, that is of the order of the roughness measured for instance for Nylon particle surface \cite{Gondret2002}, but still below the range of $\delta z/R$ observed in the experiments. Only values of $\gamma$ larger than $10$ could allow for a better match. Hence the apparent roughness is more likely to relate to the material roughness, which is identical for all the cylinders (outer shell in PVC bouncing on a PVC plate).

Nevertheless, the rigid body approximation is clearly an upper bound for the evolution of $e$ (numeric) and $e_y/e_y^0$ (experiment) as a function of $St/St_c$. This observation suggests that the elasticity parameter $\alpha$ used in the numerical model, and its apparent counterpart in experiments, has to be considered. We recall here that by apparent contact elasticity, we refer to the fact that the contact time occurring during this overlap scale is finite.

Figures \ref{fig:St_expe_model}(b) \& (d) display the contact time measured from numerical simulations (the time where the separation gap is smaller than $\eta R$) and the apparent contact time from the experiments, both being referred to as $t_C$, as a function of $St/St_c$, respectively. The contact time is scaled by the settling time scale $D/V_t$. These figures suggest that i) the dimensionless contact time obtained from the simulations (b) varies significantly with the elasticity parameter $\alpha$ but more weakly with the Stokes number, and that ii) it falls in the same range of contact time measured experimentally (d). The latter can be grouped in two families. One group can be described by constant $t_{C1} V_t/D=(6 \pm 3)\times10^{-2}$, corresponding mainly to the collisions with 2-contacts which lead to a large apparent contact at high Stokes numbers (filled red symbols in figure \ref{fig:St_expe_model}b). As for the other group (empty red symbols in \ref{fig:St_expe_model}b), the contact time seems to be adequately described by a law of the form $t_{C2} V_t/D = (8\pm4)\times10^{-3}\sqrt{St/St_c-1}$. Note that the empty symbols correspond to 1-contact rebounds and cavitation has been observed for the cylinders of type C (squares) and D (triangles), even in the low $St/St_c$ range.  
In the first group where the dimensionless contact time can be considered as constant (and relatively large), the contact time is close to, but exceeds the largest contact time measured in the simulations at large $\alpha$ (soft contact). This observation agrees with the experimental coefficient of restitution $e_y/e_y^{0}$ being globally smaller than the numerical value $e$ at a given $St/St_c$. However in the second group, the contact time increases by more than one order of magnitude in the considered range of $St/St_c$, in a range which is consistent with the influence of the elasticity parameter $\alpha$ on the contact time obtained from numerical simulations (b).

According to the previous observations, we shall therefore discuss in the following the possibility of modelling the coefficient of restitution given a law for the contact time, whatever its physical origin (here elasticity in simulations and more complex processes as cavitation, orientation and multi-contact in experiments). 
From eqs. \eqref{eq:Vc_Vt} and \eqref{eq:Vr_Vc0}, if one uses the relation between cut-off length $\eta$, the critical Stokes number following eq. (\ref{eq:Stc}) and the definition of $St$ in \eqref{eq:St}, one obtains
\begin{equation}
\label{eq:modelrestitution}
e_{mC} =  \left(1 - \frac{St_c}{St}\right) \exp \left[ -\frac{\pi }{3\sqrt{2}}\left(\frac{ St_c}{\gamma}\right)^3 \frac{1}{\tilde{S}\, St}\frac{t_c V_t}{D}\right]
\end{equation}
where $\tilde{S} = S_{obj}/D^2$ with $S_{obj}$ the 2D surface of the falling object. For a disc, $S_{obj} = \pi D^2/4$ while for the experimental cylinder with a tail, one has $S_{obj} = \pi D^2/4 + S_{tail}\approx 2D^2$ for the geometry of tail used here. For numerical simulations, considering a constant contact time for a given elastic parameter (lines in figure \ref{fig:St_expe_model}(b)), model \eqref{eq:modelrestitution} leads to the effective coefficient of restitution shown by solid lines in figure \ref{fig:St_expe_model}(a), with colors corresponding to the ones used for the contact time. Note that such results is closly related to the one shown in figure \ref{fig:St_e_num_model}(b). Even if this model does not perfectly collapse onto the numerical results of $e$, it clearly highlight the right quantitative trend.   

Let's now consider the experimental results. Firstly, data corresponding to a constant contact time $t_{C1} V_t/D=(6 \pm 3)\times10^{-2}$ are discussed. Following eq. (\ref{eq:modelrestitution}), this contact time law leads to
\begin{equation}
\label{eq:e_model1}
e_{mC1} \simeq \left(1 - \frac{St_c}{St}\right) \exp \left[ -\frac{\pi }{6\sqrt{2}}\left(\frac{ St_c}{\gamma}\right)^3 \frac{(6\pm3)\,10^{-2}}{St}\right]
\end{equation}
The green solid lines in figure \ref{fig:St_expe_model}(c,d) describing this model, with $St/St_c=75$ and $\gamma=2$, offer a good description of the experimental results for values of $St/St_c$ large enough, with a strong mismatch for values between $1$ and $20$. Note moreover that such contact law mimic the one used previously to characterize the influence of the elasticity on numerical results. Similar trends are clearly observed, but here for a 'softer' experimental contact (large contact time).
Lastly, accounting for a variations of this apparent contact time with the Stokes number (cyan in figure \ref{fig:St_expe_model}(d)) leads to a different law. 
In particular, using $t_{C2} V_t/D = (8\pm4)\,10^{-3}\sqrt{St/St_c-1}$, implies for the coefficient of restitution based on \eqref{eq:modelrestitution}, as
\begin{equation}
\label{eq:e_model2}
e_{mC2} \simeq \left(1 - \frac{St_c}{St}\right) \exp \left[ -\frac{\pi }{6\sqrt{2}}\left(\frac{ St_c}{\gamma}\right)^3 \frac{(8\pm4)\,10^{-3}\sqrt{St/St_c-1}}{St}\right]\,.
\end{equation}
Once again, we consider $St_c=75$ and $\gamma=2$ for experiments. The cyan line in figure \ref{fig:St_expe_model}(b) corresponds to the coefficient of restitution $e_{mC2}$. Note that this law provide a better prediction of the experimental data at relatively small $St/St_c$, mostly open symbols in our case. To finish with, the shaded areas in figure \ref{fig:St_expe_model}(c,d) indicate the influence of the uncertainties in estimating the fitting parameters for a value of $St_c=75$. In appendix \ref{sec:app_Stceffect}, figures \ref{fig:ey_model_compare}(a) and (b) show the same model with different $St_c$, keeping $\gamma=2$. The values $50$ and $100$ were used respectively in panels (a) and (b). Comparison with figure \ref{fig:St_expe_model}(a) suggests that using $St_c=75$ leads to a better matching between the model and our data.   
In summary, these models allow to describe the features of the rebound in two different limits: the low $St/St_c$ limit where 1-contact with relatively short dimensionless time and sometimes cavitation are observed, and the high $St/St_c$ where rather collisions with 2-contact occur due to cylinder pitching.

Before ending this section, we test here a simple scaling of the solid contact time following Izard et al. \cite{izard2014modelling}. For this purpose, if one assumes a balance between the force of elastic deformation that scales like $k_n\eta$ and the lubrication force that scales like $\lambda |V_c|$, the ratio $V_r/V_c$ can be written in a way that does not depend on the cylinder rigidity, but only on particle inertia and roughness which sets the critical Stokes number $St_c$. After some calculation summarized in appendix \ref{annexe:tc_eIz}, we obtain the velocity ratio $\frac{V_r}{V_c} = - exp \left[ - \frac{\pi}{2\sqrt{2}} \sqrt{\frac{St_c}{St-St_c}}\right]$. The corresponding restitution coefficient (called $e_{Iz}$) is thus 
\begin{equation}
    e_{Iz} = \left( 1-\frac{St_c}{St} \right) exp \left[ - \frac{\pi}{2\sqrt{2}} \sqrt{\frac{St_c}{St-St_c}}\right] \,.
    \label{eq:e_Iz}
\end{equation}
and the associated contact time (called here $\tau_{Iz}$) is $\frac{\tau_{Iz} V_t}{D} = \frac{\pi}{2}\left(\frac{3}{\sqrt{2}}\frac{\rho_p}{\rho^*}\frac{St}{1-St_c/St}\right)^{1/2} \eta^{5/4}$. 
The difference with the models above lies in the constant elasticity. This model is compared against numerical simulations and experiments (blue lines in figure \ref{fig:St_expe_model}). It under predicts the measured  contact time, and in a consistent way, the coefficient of restitution obtained from this model (blue line in fig. \ref{fig:St_expe_model}a) follows the same S-shape but over predicts the experimental results. Thus, while the simple model of \cite{izard2014modelling} captures qualitatively the features of the rebound and is closer to the experimental results than the rigid-body model, it fails to predict the wide range of experimental $e-St$ data that seems to be associated with more complex apparent elasticity during contact. The latter ingredient is therefore essential to allow a general description of the rebound of non-spherical objects.

\section{Conclusion}\label{sec:conc}

To conclude this work, we have shown through figure \ref{fig:St_expe_model} that using a critical Stokes number $St_c$ and a relevant contact time $t_c$ allows to capture the characteristics of bouncing, especially the coefficient of restitution. In an idealized 2D configuration, $St_c$ and $t_c$ were associated with a rebound cut-off length $\eta$ and an elasticity parameter $\alpha$, with in particular $t_c\equiv \tau$ emanating from a a visco-elastic model, whereas in the experiments, $St_c$ and $t_C$ were inferred from measurements (image processing).
Even if a perfect matching is not obtained, the evolution of the coefficient of restitution and contact time scale with the Stokes number and their range of dispersion are quite consistent in view of the difficulty to measure such quantities in experiments. Note again that mechanisms at play during experimental contact are numerous: pitching leading  to multiple contact, cavitation, among other, for which the implementation of a contact elasticity in models remain an empirical parameterization. A finer description requires larger database, which is left to future investigation.

%%%%%%%%%%%%%%%%%%%%%%%%%%%%%%%%%%%%%%%%%%%%%%%%%%%%%%%%%%%%%%%%%%%%%%%%%%%%%
\section{Acknowledgments}

Computational resources were provided by the computing meso-centre CALMIP under project no. P1002. The research federation FERMaT is acknowledged for the optical measurement facilities. AAC thanks CONACYT AND CIC-UMSNH for their financial support.

%%%%%%%%%%%%%%%%%%%%%%%%%%%%%%%%%%%%%%%%%%%%%%%%%%%%%%%%%%%%%%%%%%%%%%%%%%%%%

\section{Appendix}
\subsection{Lubrication force between a cylinder and a wall}
\label{sec:app_Flub}

\begin{figure}
  \centering
  \includegraphics[trim=0 0 0 100, scale=0.5]{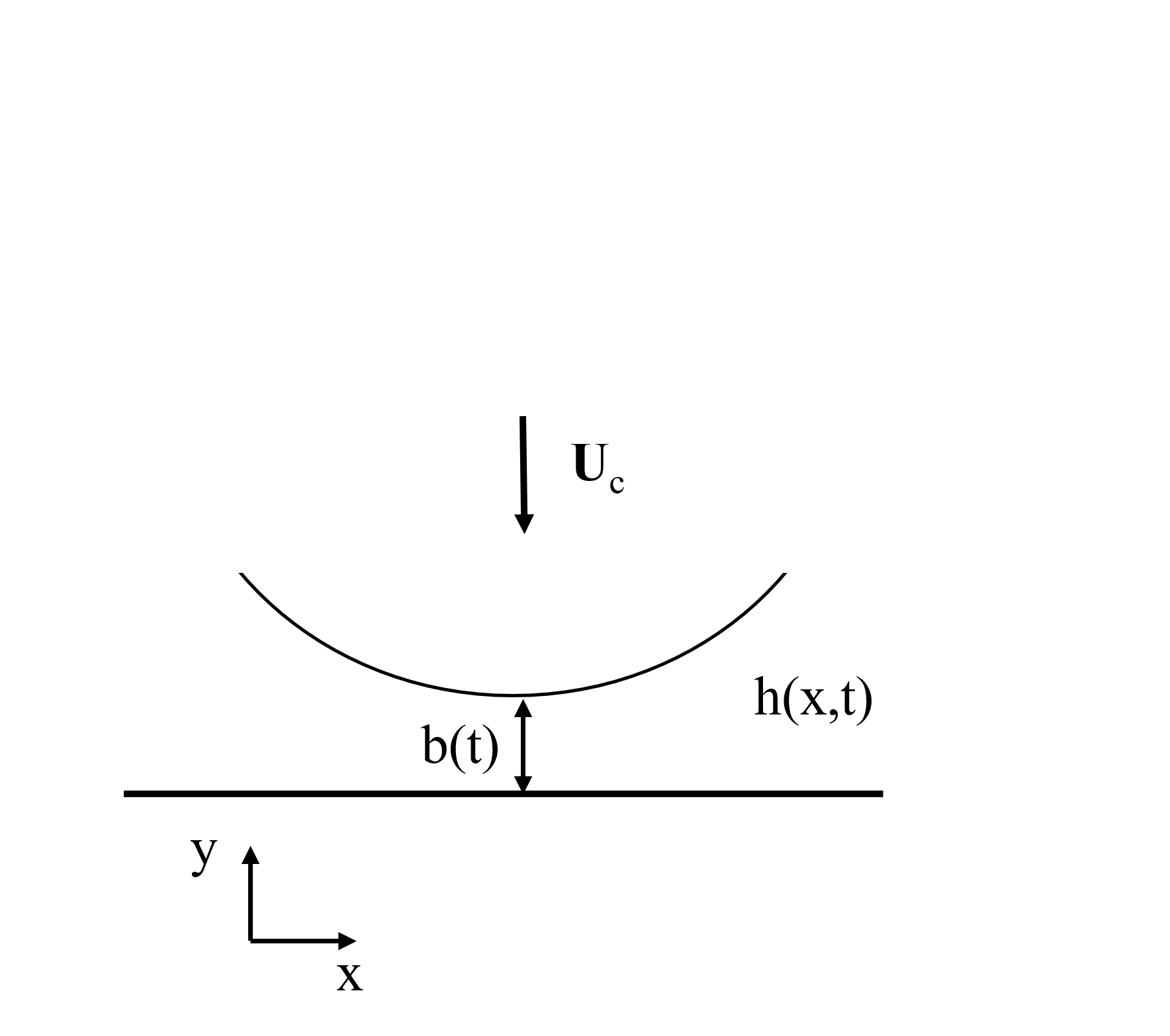}
  \caption{Configuration used for the calculation of the viscous lubrication force in the gap between an infinite circular cylinder (of radius R) and a rigid plane.} 
\label{fig:geometry_lubrication}
\end{figure}

Eq. \ref{eq:Flub} was obtained in the frame of the thin lubrication approximation \cite{leal2007advanced}. The cylinder is assumed to approach the wall with a given normal velocity $\textbf{U}_c$. The coordinates $x$ and $y$ are equal to 0 at the center line at the wall (see the geometry in figure \ref{fig:geometry_lubrication}). We assume that the problem is quasi-steady (the time variation of the flow velocity $\textbf{u}$ and pressure $p$ is small compared to their spatial variation). The separation gap $h$ varies in time due to the cylinder motion according $\frac{\partial h}{\partial t}=U_{c,y}$, where we recall that the separation gap is not uniform along x, i.e. $h(x,t)=b(t)+\frac{R}{2}\frac{x^2}{R^2}$ at the leading order, with $b$ denoting the minimum height. In the frame of the thin gap lubrication assumptions, the flow in the gap is dominantly along the $x$ direction, while the pressure does not vary significantly along the $y$ direction. After finding the flow velocity profile and assuming a constant flow rate along $x$, one obtains the following relation between the gap $h$ and the pressure $p$.

\begin{equation}
    \frac{\partial h}{\partial t} = \frac{d}{dx} \left( \frac{h^3}{12\mu} \frac{dp}{dx}\right)
\end{equation}

\noindent This leads to a differential equation for the pressure, which after integration along $x$ becomes: 

\begin{equation}
    p = 12\mu \frac{\partial h}{\partial t} \int_0^x  \frac{x}{\left(b+\frac{x^2}{2R}\right)^3}   dx
\end{equation}

\noindent We assume that at a certain distance $R_0$ from the center line ($0<R_0<R$), the pressure tends toward the outer pressure $p\rightarrow p_0$ and the lubrication assumptions do not hold anymore. The pressure profile along $x$ direction becomes then:

\begin{equation}
    p(x)-p_0 = 12\mu \frac{\partial h}{\partial t} \left[ \frac{R}{2\left( b+\frac{R_0^2}{2R}\right)^2} -  \frac{R}{2\left( b+\frac{x^2}{2R}\right)^2}   \right]
\end{equation}

\noindent The wall-normal lubrication force can be obtained by integrating the pressure along the $x$ direction, and assuming small gaps $\epsilon = \frac{b}{R}\rightarrow0$:

\begin{equation}
    F_y = -12\pi\mu \frac{\partial h}{\partial t} \left(\frac{1}{2\epsilon}\right)^{3/2}
\end{equation}

\subsection{Calculation of the collision-to-terminal velocity ratio}
\label{annexe:vc_vt}

\noindent Integration of \eqref{eq:motion}-\eqref{eq:Flub} from a distance where the cylinder starts to decelerate ($V\approx V_t$ at $\zeta\approx R$) to the wall where collision occurs ($V=V_c$ at $\zeta= \eta R$) leads to $
m^* \int_{V_t}^{V_c} dV = - \frac{6}{\sqrt{2}} \pi \mu  R^{3/2} \int_{R}^{\eta R} \frac{d\zeta}{\zeta^{3/2}}$. As the roughness $\eta R$ is very small compared to the cylinder radius $R$ the equality then approximates to $m^*(V_c - V_t) \simeq \frac{12}{\sqrt{2}} \pi \mu R \left(\frac{1}{\eta}\right)^{1/2}$, or similarly, $\frac{V_c}{V_t} \simeq 1 - \frac{8}{3\sqrt{2}} \frac{1}{St} \left(\frac{1}{\eta}\right)^{1/2}$. 
At this stage, we can define the critical Stokes number, that corresponds to the minimum inertia required for the cylinder to reach the wall with a finite velocity that allows the rebound. This critical Stokes number corresponds to $V_c=0^+$, which leads to $St_c = \frac{8}{3\sqrt{2}} \left(\frac{1}{\eta}\right)^{1/2}$. Therefore the collision-to-terminal velocity ratio is given by: 

\begin{equation}
\frac{V_c}{V_t} = 1 - \frac{St_c}{St}.
\end{equation}

\subsection{Estimation of the contact time from simple scaling analysis}
\label{annexe:tc_eIz}

Above the rebound onset, the contact time corresponding to the damped oscillator harmonic can be approximated by $\tau=\pi \sqrt{\frac{m^*}{k_n}}$, as explained below eq. \ref{eq:Vr_Vc0}. Yet, the assumption of equilibrium between elastic and viscous lubrication forces during the contact stage leads to $\lambda |V_c| \approx k_n \eta R$. Here, $\lambda$ corresponds to the coefficient of the lubrication force as defined in \ref{eq:damp_oscill}, the distance between the particle surface and the wall is approximated by $\eta R$. After replacing the contact velocity $V_c$ from eq. \ref{eq:Vc_Vt}, one can estimate the ratio $m^*/k_n$, and thus obtain the contact time, that we will call here $\tau_{Iz}$:

\begin{equation}
\frac{\tau_{Iz} V_t}{D} = \frac{\pi}{2}\left(\frac{3}{\sqrt{2}}\frac{\rho_p}{\rho^*}\frac{St}{1-St_c/St}\right)^{1/2} \eta^{5/4} =\frac{\pi}{2\Gamma} \eta^{5/4}\,, \label{eq:tc_Iz}
\end{equation}
with $1/\Gamma = \left(3\rho_p St / \sqrt{2} \rho^* (1 -St_c /St)\right)^{1/2}$

\subsection{Influence of $St_c$ on $e_{mC}$ }
\label{sec:app_Stceffect}

\begin{figure}
  \centering
  (a) \hspace{8cm} (b)\\  
  \includegraphics[height=0.25\textheight]{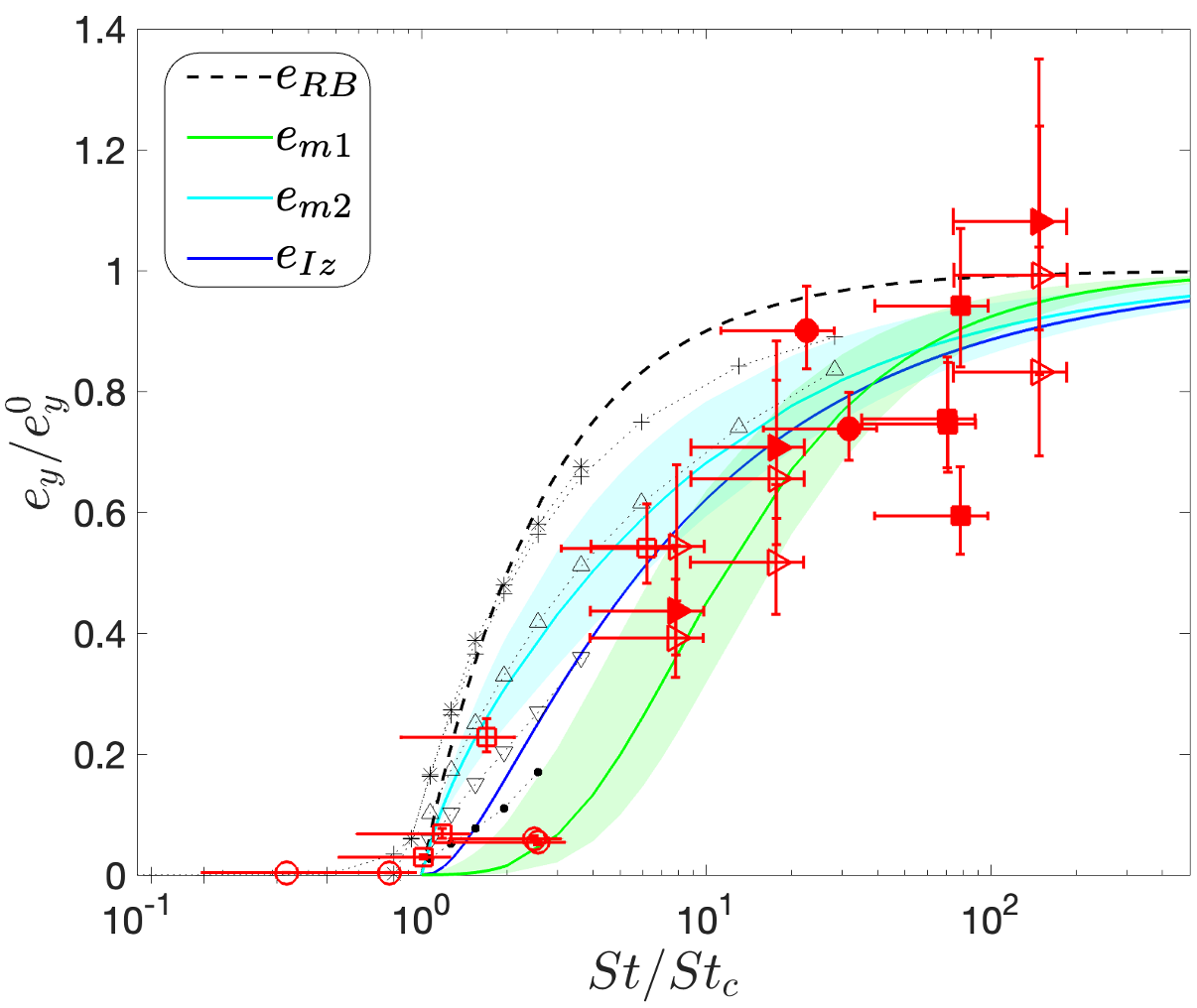}
    \includegraphics[height=0.25\textheight]{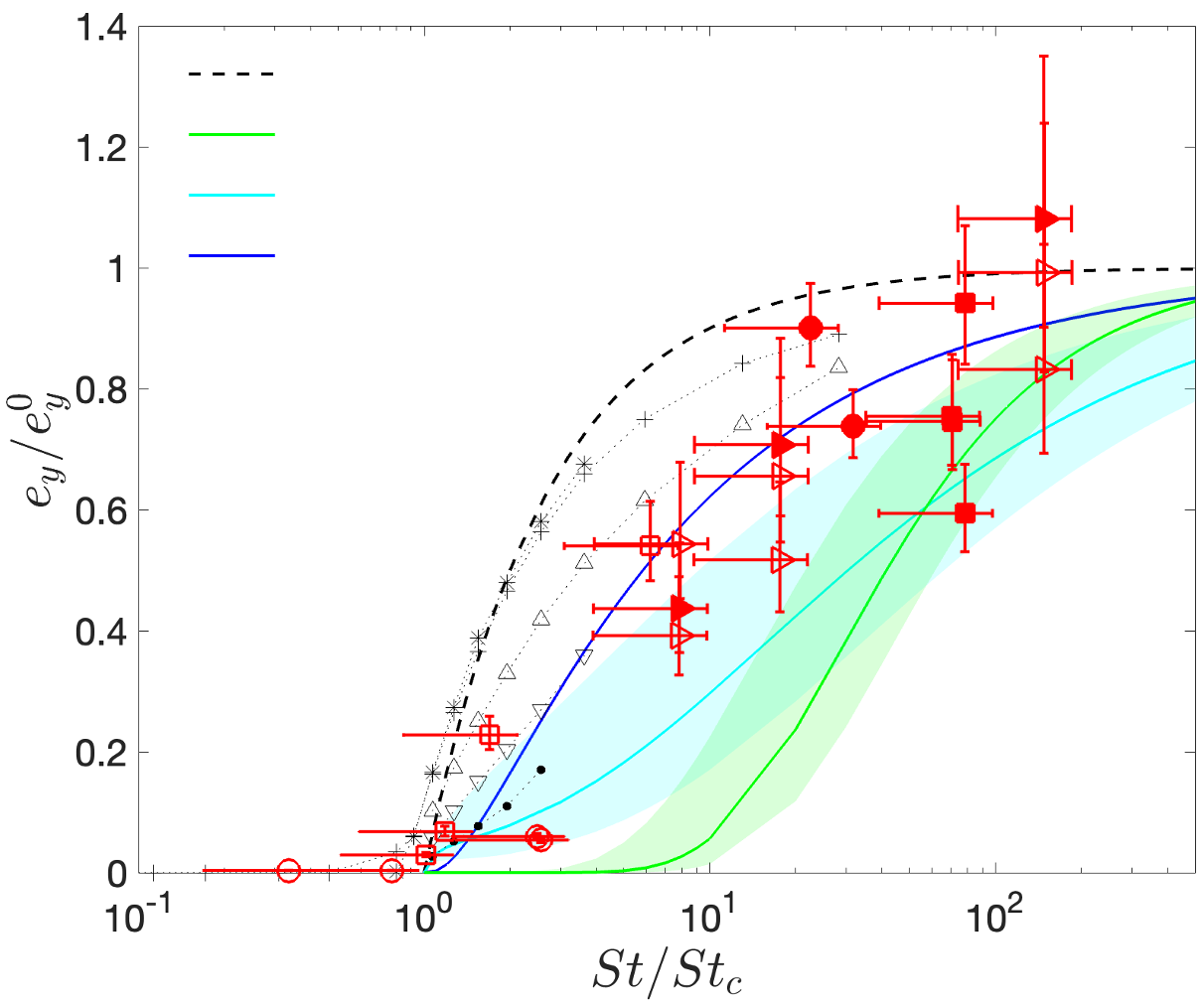}
  \caption{Same as in Fig.\ref{fig:St_expe_model}(a) but with different values of the threshold for rebounds in models $e_{m1}$ and $e_{m2}$ with $\gamma = 2$, (a) $St_c=50$ and (b) $St_c=100$.}
\label{fig:ey_model_compare}
\end{figure}

In Figure \ref{fig:ey_model_compare}, we compare the influence of the value of $St_c$ on the models for the coefficient of restitution, $e_{mC1}$ and $e_{mC2}$, described in equations \eqref{eq:e_model1} and \eqref{eq:e_model2} respectively.

%%%%%%%%%%%%%%%%%%%%%%%%%%%%%%%%%%%%%%%%%%%%%%%%%%%%%%%%%

%\bibliography{Biblio}% Produces the bibliography via BibTeX.

%\begin{thebibliography}{10}

%\end{thebibliography}

\end{document}